\documentclass[twocolumn, twocolappendix]{aastex701}

\usepackage{graphicx}
\usepackage{hyperref}

\usepackage{amsmath}

\begin{document}

\title{JWST-MEP: Limb Asymmetries and Possible Evidence for Time Variability in Kepler-12 b}

\shorttitle{Limb Asymmetries and Possible Evidence for Time Variability in Kepler-12 $\mathrm{b}$}

\author[orcid=0000-0002-0931-735X]{Måns Holmberg}
\affiliation{Space Telescope Science Institute, 3700 San Martin Drive, Baltimore, MD 21218, USA}
\email[show]{mholmberg@stsci.edu}  

\author[orcid=0000-0001-9513-1449]{Néstor Espinoza} 
\affiliation{Space Telescope Science Institute, 3700 San Martin Drive, Baltimore, MD 21218, USA}
\email{nespinoza@stsci.edu}

\author[0009-0002-8461-6111]{Yanbo Pan} 
\affiliation{Department of Astronomy, University of Virginia, 530 McCormick Road, Charlottesville, VA 22904, USA}
\email{war8sk@virginia.edu}

\author[0000-0003-4816-3469]{Ryan J. MacDonald} 
\affiliation{School of Physics and Astronomy, University of St Andrews, North Haugh, St Andrews, KY16 9SS, UK}
\email{Ryan.MacDonald@st-andrews.ac.uk}

\author[orcid=0000-0001-8342-1895, gname='Maria', sname='Steinrueck']{Maria E. Steinrueck}
\altaffiliation{51 Pegasi b fellow}
\affiliation{Department of Astronomy \& Astrophysics, University of Chicago, Chicago, IL 60637, USA}
\email{msteinrueck@uchicago.edu}  

\author[]{Thomas Kennedy} 
\affiliation{Department of Astronomy, University of Michigan, 1085 S University Ave, Ann Arbor, MI 48109, USA}
\email{thomak@umich.edu}

\author[0000-0002-8956-2047]{Dominic Samra} 
\affiliation{Department of Astronomy \& Astrophysics, University of Chicago, Chicago, IL 60637, USA}
\email{dominicsamra.uk@gmail.com}

\author[0000-0003-0156-4564]{Luis Welbanks}
\affiliation{School of Earth and Space Exploration, Arizona State University, 781 Terrace Mall, Tempe, AZ 85287, USA}
\email{luis.welbanks@asu.edu}

\author[0000-0002-9158-7315]{Rafael Brahm}
\affiliation{Facultad de Ingeniería y Ciencias, Universidad Adolfo Ibáñez, Avenida Diagonal Las Torres 2640, 7941169 Peñalolén, Santiago, Chile}
\email{rafael.brahm@uai.cl}

\author[]{Eva-Maria Ahrer}
\affiliation{Max-Planck-Institut f\"{u}r Astronomie, K\"{o}nigstuhl 17, 69117 Heidelberg, Germany}
\email{ahrer@mpia.de}

\author[orcid=0000-0002-4997-0847]{Duncan A. Christie}
\affiliation{Max Planck Institute for Astronomy, K\"onigstuhl 17, D-69117 Heidelberg, Germany}
\email{christie@mpia.de}

\author[]{Wolf Cukier}
\affiliation{Department of Astronomy \& Astrophysics, the University of Chicago, Chicago, IL, 60637, USA}
\email{wcukier@uchicago.edu}

\author[]{Carlos Gascón}
\affiliation{Space Telescope Science Institute, 3700 San Martin Drive, Baltimore, MD 21218, USA}
\email{cgascon@stsci.edu}

\author[orcid=0000-0003-1907-5910]{Kevin Heng}
\affiliation{Ludwig Maximilian University, Faculty of Physics, Scheinerstr. 1, Munich D-81679, Germany}
\affiliation{Munich Center for Geoastronomy, Ludwig Maximilian University, Theresienstrasse 41, D-80333, Munich, Germany}
\affiliation{University College London, Department of Physics \& Astronomy, Gower St, London, WC1E 6BT, United Kingdom}
\email{Kevin.Heng@physik.lmu.de}

\author[orcid=0000-0001-6707-4563]{Nathan J. Mayne}
\affiliation{Department of Physics and Astronomy, University of Exeter, Exeter EX4 4QL, UK}
\email{N.J.Mayne@exeter.ac.uk}

\author[orcid=0000-0001-8342-1895]{Matthew C. Nixon}
\altaffiliation{51 Pegasi b fellow}
\affiliation{School of Earth and Space Exploration, Arizona State University, Tempe, AZ, USA}
\email{matthewnixon@asu.edu}

\author[]{Diana Powell}
\affiliation{Department of Astronomy \& Astrophysics, University of Chicago, Chicago, IL 60637, USA}
\email{diana.powell@uchicago.edu}

\author[orcid=0000-0003-3963-9672]{Emily Rauscher}
\affiliation{Department of Astronomy, University of Michigan, 1085 S University Ave, Ann Arbor, MI 48109, USA}
\email{erausche@umich.edu}

\author[]{Yoav Rotman}
\affiliation{School of Earth and Space Exploration, Arizona State University, 781 Terrace Mall, Tempe, AZ 85287, USA}
\email{yrotman@asu.edu}

\author[]{Arjun B. Savel}
\affiliation{Astronomy Department, University of Maryland, College Park, 4296 Stadium Dr., College Park, MD 20742 USA}
\email{asavel@umd.edu}


\begin{abstract}

The quality of JWST observations has opened a new frontier in exoplanet atmospheric characterisation, enabling transmission spectroscopy to directly resolve the terminator and probe atmospheric dynamics. We present two JWST/NIRSpec transmission spectra of the hot Jupiter Kepler-12~b that resolve the planet's morning and evening limbs. We identify pronounced limb asymmetries, with temperatures at 1~mbar of $1546_{-63}^{+65}$~K and $900_{-100}^{+110}$~K on the evening and morning limbs, respectively, corresponding to a temperature contrast of $650_{-140}^{+130}$~K. The morning limb is also substantially cloudier, while the evening limb shows no evidence for aerosols in the observable atmosphere. This temperature contrast is markedly larger than predicted by general circulation models ($\lesssim300$--$400$~K) and exceeds previous measurements for other exoplanets ($\lesssim450$~K). We detect H$_2$O ($\ln B = 350.9$) and CO$_2$ ($\ln B = 8.2$), and find evidence for H$_2$S ($\ln B = 4.1$), and infer subsolar elemental abundances. Despite the strong thermal and aerosol contrasts, we find no statistically significant differences in gas-phase abundances between the limbs, although a higher CO$_2$ abundance on the morning limb is weakly favoured. In addition to spatial structure, we detect significant visit-to-visit variability, although its physical origin remains more uncertain: it could arise from changes in the planetary atmosphere, such as cloud variability, or from potential stellar variability. We further show that properly accounting for inter-wavelength and inter-limb covariance is essential for limb-resolved spectroscopy, yielding substantially improved constraints on atmospheric properties. Together, these results demonstrate JWST transmission spectroscopy as a robust probe of thermal asymmetries and aerosol distributions in giant exoplanet atmospheres.

\end{abstract}



\section{Introduction}

\label{sec:intro}

Atmospheric circulation plays a fundamental role in shaping the observable properties of exoplanet atmospheres. For close-in, tidally locked planets, intense stellar irradiation drives large day--night temperature contrasts, giving rise to complex three-dimensional atmospheric structures and dynamical processes \citep[e.g.,][]{showman2009}. As a result, the planetary terminator is not expected to be homogeneous, but instead to reflect distinct physical and chemical conditions across the morning and evening limbs \citep[e.g.,][]{Fortney2010}. With the advent of the James Webb Space Telescope (JWST), we now have the precision and stability required to directly probe these effects \citep[as first demonstrated by][]{Espinoza2024, Murphy2024b}, opening a new window into the dynamics and chemistry of exoplanet atmospheres. General circulation models (GCMs) and cloud microphysics models predict substantial differences in temperature, chemistry, and cloud properties across the terminator, driven by the interplay of atmospheric advection, microphysical processes, and radiative timescales \citep[e.g.,][]{Kataria2016, Powell2019, Roman2021, Zamyatina2024, Mak2025}.

Observational evidence for atmospheric asymmetries has steadily grown in recent years. Ground-based high-resolution spectroscopy has revealed asymmetries in atmospheric winds and composition for several highly irradiated exoplanets \citep[e.g.,][]{Louden2015, Bourrier2020, Hoeijmakers2020, Ehrenreich2020, Kesseli2021, Gandhi2022}. More recently, limb-resolved analyses with JWST have begun to directly separate the transmission spectra of the morning and evening hemispheres. These studies have reported differences in temperature structure, cloud coverage, and molecular abundances across a range of planets \citep[e.g., WASP-39~b, WASP-107~b, and WASP-94~A~b;][]{Espinoza2024, Murphy2024b, Mukherjee2026, Fu2025}, demonstrating the potential of JWST to spatially resolve exoplanet atmospheres during transit.

Despite these advances, robustly characterizing limb asymmetries remains challenging. The inferred morning and evening spectra are strongly degenerate with each other and with key system parameters, including transit timing and orbital geometry \citep{espinoza_limb_2015, Espinoza2024, Murphy2024a}. These degeneracies induce significant covariance between the spectra of the two limbs \citep{espinoza_limb_2015}. If neglected, this covariance can lead to loss of information and potentially biased atmospheric inferences. Addressing this challenge requires improved analysis techniques. In this work, we introduce a framework that explicitly accounts for inter-limb and inter-wavelength covariance, building on the formalism developed by \cite{espinoza_limb_2015}.

In addition to spatial inhomogeneities, exoplanet atmospheres also exhibit temporal variability. Time-dependent processes such as evolving cloud coverage, dynamical instabilities, and radiative feedback may produce changes in atmospheric temperature structure and composition \citep[e.g.,][]{Parmentier2013, Lines2018, Komacek2020}. While variability is well established in brown dwarfs and directly imaged exoplanets \citep[see the review by][]{Biller2017}, evidence for time-variable atmospheres in transiting giant exoplanets remains limited, with only tentative evidence reported to date \citep{Armstrong2016, Jackson2019, Bell2019, Asnodkar2022, Changeat2024}. Establishing the presence and origin of such variability is therefore a major objective for JWST, particularly for multi-epoch observations that are commonly analysed under the assumption of a steady state atmosphere.

In this work, we present two JWST/NIRSpec PRISM observations of the giant exoplanet Kepler-12~b \citep{Fortney2011} obtained as part of the JWST Morning/Evening Program (MEP; JWST GO program 3969; PIs: N.~Espinoza \& D.~Powell). The MEP survey targets seven hot Jupiters spanning equilibrium temperatures of approximately $1000$-$2000$~K with broadly similar bulk properties, enabling comparative studies of atmospheric asymmetries across a controlled sample. As part of this survey, we perform the first atmospheric characterisation of Kepler-12~b using limb-resolved transmission spectroscopy. With a radius of R$_\mathrm{p} = 1.7~\mathrm{R}_\mathrm{J}$, a mass of M$_\mathrm{p} = 0.43~\mathrm{M}_\mathrm{J}$, and an equilibrium temperature of T$_\mathrm{eq} \approx 1500$~K (A$_\mathrm{B} = 0$), Kepler-12~b is a highly inflated hot Jupiter with a large atmospheric scale height. Its optical phase curve exhibits a westward-shifted brightness peak indicative of asymmetric cloud coverage \citep{Esteves2015, Angerhausen2015}, making it an especially compelling target for limb-resolved studies with JWST. Moreover, the host star appears photometrically and chromospherically quiet  \citep{Fortney2011, Esteves2015}, minimizing the impact of stellar contamination and variability. The two visits presented here are separated by approximately two months, enabling a joint investigation of both spatial structure and temporal variability in the planet’s atmosphere.

This paper is organised as follows. In Section~\ref{sec:obs}, we describe the observations and data reduction. In Section~\ref{sec:retrievals}, we present the atmospheric retrieval framework and results. In Section~\ref{sec:discussion}, we discuss the implications of our findings and summarise our conclusions.

\section{Observations and Data Reduction} 
\label{sec:obs}

We observed two transits of the giant exoplanet Kepler-12~b using JWST/NIRSpec PRISM on 22 April 2024 and 2 July 2024, covering the near-infrared wavelength range from 0.6--5.4~$\mu$m. Each observation consists of 29,131 integrations with 5 groups per integration, corresponding to a total duration of approximately 11~hours per visit. Given the target's magnitude of $J = 12.4$, the observations remain below the detector saturation limit for all wavelengths. During the first visit, the high-gain antenna (HGA) moved during the transit at a phase of around 0.01. However, this event produced no measurable long-term effects in the light curve, and we simply masked the affected integrations. The pointing stability of both visits is nominal for JWST observations, with an average standard deviation of $0.001$ pixels after binning every 10 integrations (14-second cadence). Overall, both datasets are of high quality, with no evidence for significant instrumental systematics.

After reducing the data and extracting the light curves, we first constrain the system parameters using the white-light curves. We then generate transmission spectra at native spectral resolution (binning every two detector pixels) and estimate the associated covariance matrices using a new analysis framework introduced in this work. Additionally, we characterise correlated-noise properties in Appendix~\ref{app:noise}, limb darkening properties in Appendix~\ref{app:limb-darkening}, and light-curve trends in Appendix~\ref{app:trend}, which we use to generate the transmission spectra.

\subsection{Data reduction}

\begin{figure*}
\centering
	\includegraphics[width=0.99\textwidth]{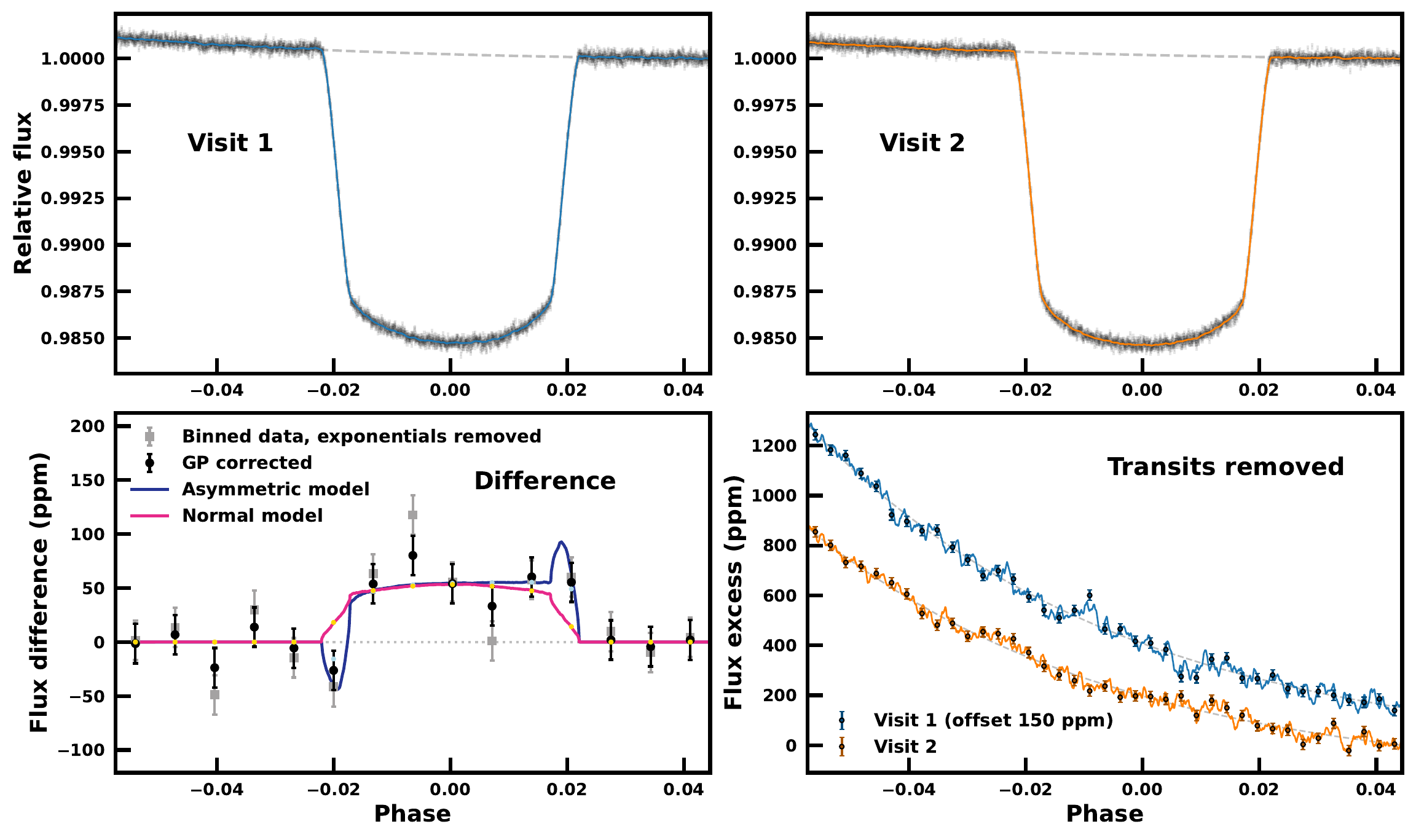}
    \vspace{-2mm}
    \caption{{\bf Top:} The Kepler-12~b white-light curves of both visits. The blue and orange curves represent the median models, while the dashed grey curves represent the out-of-transit model flux. {\bf Bottom:} The left panel shows the difference between the two white-light curves. The grey data points represent the raw data, after removing the exponential trends, while the black data points show the data after removing the exponential and GP components. These data points have been binned for clarity. The dark blue and pink model fits represent the transit models with and without limb asymmetry, respectively. The binned model points are shown in yellow and light blue, respectively. The right panel shows the light curves after dividing out the transit model to illustrate the exponential trends and time-correlated noise.
    }
    \label{fig:wlc} 
\end{figure*}

We adopt \texttt{JExoRES} \citep{holmberg2023, Madhu_carbon2023} to reduce the observations. This involves using the JWST Science Calibration Pipeline \citep{Bushouse2020} for Stages 1-2 to calibrate the data. During Stage 1, we perform an additional column-by-column background subtraction step to mitigate 1/f noise at the group level. In Stage 3, we then extracted the spectra using the optimal extraction algorithm \citep{horne_optimal_1986}, while iteratively rejecting $5\sigma$ outliers. For the extraction, we used the first three principal components of the data cube as a time-varying model of the point spread functions (PSFs). This analysis did not reveal any sudden morphological changes to the PSFs. For each integration, we mask spectral channels with more than 20\% of their flux removed due to outliers. We then perform sigma clipping on the extracted light curve of each spectral channel using a $4\sigma$ threshold, where the transit signal was removed by a median high-pass filter. 

Furthermore, as mentioned above, we masked 10 integrations following the HGA movement during the first visit. We also removed the first 15 minutes of each visit to exclude any short-lived settling behaviours. Finally, we binned the light curves every 10 integrations to reduce data volume, resulting in a 14-second cadence. As a robustness check, we find that all parameters from the white light curve fit remain consistent, as well as the transmission spectra themselves, regardless of whether we bin the light curves over time.
 
\subsection{White-light light-curve analysis} \label{sec:orbital_properties}

We jointly fit the band-integrated (0.6-5.4~$\mu$m) white-light curves of the two visits to refine the orbital parameters of Kepler-12~b. We model the transit light curves using \texttt{batman} \citep{kreidberg_batman_2015} and \texttt{catwoman} \citep{Espinoza2021, Jones2022}. Prior to the analysis, we corrected an issue regarding the calculation of the geometry in \texttt{catwoman}, as described in Appendix~\ref{app:catwoman}. This fix is now implemented in the latest version of the code (version 1.1.0). To account for time-correlated noise, we employed Gaussian process (GP) models using the \texttt{celerite} package \citep{Foreman-Mackey2017}. Posterior distributions were sampled using nested sampling via \texttt{MultiNest} \citep{Feroz2009, Buchner2014} with 1000 live points.

We considered two canonical models: (1) a limb-averaged transit model using \texttt{batman}, and (2) a limb-resolved model using \texttt{catwoman}. In both cases, we adopted informative priors from \cite{Fortney2011} on the orbital period $P$, impact parameter $b$, stellar density $\rho_\star$, $\sqrt{e}\cos\omega$, and $\sqrt{e}\sin\omega$, where $e$ and $\omega$ are the eccentricity and argument of periastron. For the limb-averaged model, we additionally imposed informative priors on the mid-transit time T$_0$, while for the limb-resolved model we adopted a broader uniform prior on T$_0$. We obtain consistent system parameters when instead assuming a circular orbit and adopting broad uniform priors on all orbital parameters. Moreover, fitting separate mid-transit times for the two visits using wide uniform priors does not significantly influence the inferred limb-averaged transit depths; the corresponding mid-transit times are $60422.262046_{-0.000025}^{+0.000026}$ and $60493.269541_{-0.000026}^{+0.000025}$~days, respectively, assuming no limb asymmetry. For the limb-asymmetric model, we fix the angle between the two hemispheres to $\phi = 90^\circ$. We define the morning and evening transit depths as the effective area ratios between the semicircles and the star, such that the total transit depth satisfies $\delta = \delta_\mathrm{M} + \delta_\mathrm{E}$. For limb darkening, we employ the quadratic limb-darkening law using the parameterisation and priors from \citet{Kipping2013}. We let the limb darkening be the same for both visits, as we find no evidence for variable limb darkening coefficients.

Our canonical fits include an exponential settling trend (implemented according to Appendix~\ref{app:trend}) with a common timescale between the two visits, an exponential GP kernel with shared hyperparameters, independent white-noise jitter terms for each visit, and independent transit depths between epochs. The white-light curves are shown in Figure~\ref{fig:wlc}, while the inferred system parameters are presented in Appendix~\ref{app:system_parameters}. The standard deviations of the light curve residuals (not removing the GP components) are 183~ppm and 181~ppm for visit 1 and 2, respectively. 

The data strongly favour an exponential settling trend over a linear trend ($\ln B = 13.2$, adopting the same parameterisation as for the exponential trend, but setting $\tau \to \infty$). Allowing the exponential timescale to vary independently between visits is disfavoured ($\ln B = -2.8$), yielding e-folding timescales of $5.56_{-0.46}^{+0.56}$~hr and $6.11_{-0.68}^{+0.92}$~hr for the two visits, respectively. This is in line with recent works, which also suggest that an exponential trend with a shared timescale between exposures fit NIRSpec PRISM observations well \citep{Cassese2026, Kipping2026}. The similarity of these timescales may suggests a common origin, potentially related to detector settling, analogous to the behaviour observed with MIRI \citep[e.g.,][]{Bouwman2023, Bell2024, Fortune2025, Connors2025, Holmberg2026}.

The data strongly favour models accounting for time-correlated noise ($\ln B = 27.7$). Comparing different GP kernels, we find that an exponential kernel is moderately preferred over a stochastically driven damped harmonic oscillator (SHO) kernel with $Q = 1/\sqrt{2}$ ($\ln B = 3.0$), the latter commonly used to describe stellar granulation. However, this does not exclude stellar granulation as a possible source of the observed correlated noise.

\subsection{Evidence for visit-to-visit differences} \label{sec:white_time}

We detect a small but significant difference in the white-light transit depths between the two visits. The second visit exhibits a transit depth that is $\Delta \delta = 50_{-15}^{+16}$~ppm deeper than the first visit. When allowing for limb asymmetries, we find that most of this difference appears to originate from the evening hemisphere. The second visit for this limb is deeper by \(\Delta \delta_{\mathrm{E}} = 109_{-39}^{+38}\) ppm. In contrast, the morning hemisphere shows a smaller difference, with \(\Delta \delta_{\mathrm{M}} = -58_{-38}^{+38}\) ppm. However, the significance for a difference in the variability between the two limbs is weaker than that for the overall transit depth variation, given that $\Delta \delta_{\mathrm{E}} - \Delta \delta_{\mathrm{M}} = 167_{-76}^{+74}$~ppm corresponds to a $\sim$$2\sigma$ effect. Moreover, while the overall transit depth difference ($\Delta \delta$) and the visit-averaged limb asymmetry ($\bar \delta_{\mathrm{E}} - \bar \delta_{\mathrm{M}}$) are robust (within the uncertainties) to widening the priors on the orbital period and mid-transit time, or allowing separate mid-transit times for the two visits, the inferred values of $\Delta \delta_{\mathrm{E}}$ and $\Delta \delta_{\mathrm{M}}$ are sensitive to these assumptions. We therefore focus on the temporal variability of the limb-averaged spectrum rather than the individual limb-resolved spectra in Section~\ref{sec:time-resolved}, as the relative contributions of the morning and evening limbs to the observed variability remain uncertain.

These differences cannot be explained by inconsistencies in the inferred system parameters or by variations in stellar limb darkening, as the two visits are jointly fit and show no evidence for differing limb-darkening profiles. Letting the exponential timescales vary independently between visits does not significantly change the transit depth difference. The signal is also robust to the adopted correlated-noise model: both the canonical exponential GP kernel and the SHO kernel yield consistent results. Furthermore, the observed offset persists throughout most of the transit, rather than appearing as a localised feature, although some asymmetries are present during ingress and egress (see Figure~\ref{fig:wlc}). Explaining the signal through occulted star spots would therefore require active regions extending across most of the transit chord. Given the low impact parameter ($b = 0.18$), the absence of rotational modulation in the Kepler light curve \citep{Esteves2015}, and the low chromospheric activity inferred from HIRES spectroscopy \citep{Fortney2011}, this scenario appears unlikely. At present, we cannot distinguish between unocculted stellar heterogeneities and genuine atmospheric variability as the origin of the observed differences.

\subsection{Spectroscopic light-curve analysis and transmission spectra} \label{sec:spectrum}

\begin{figure*}
\centering
	\includegraphics[width=0.97\textwidth]{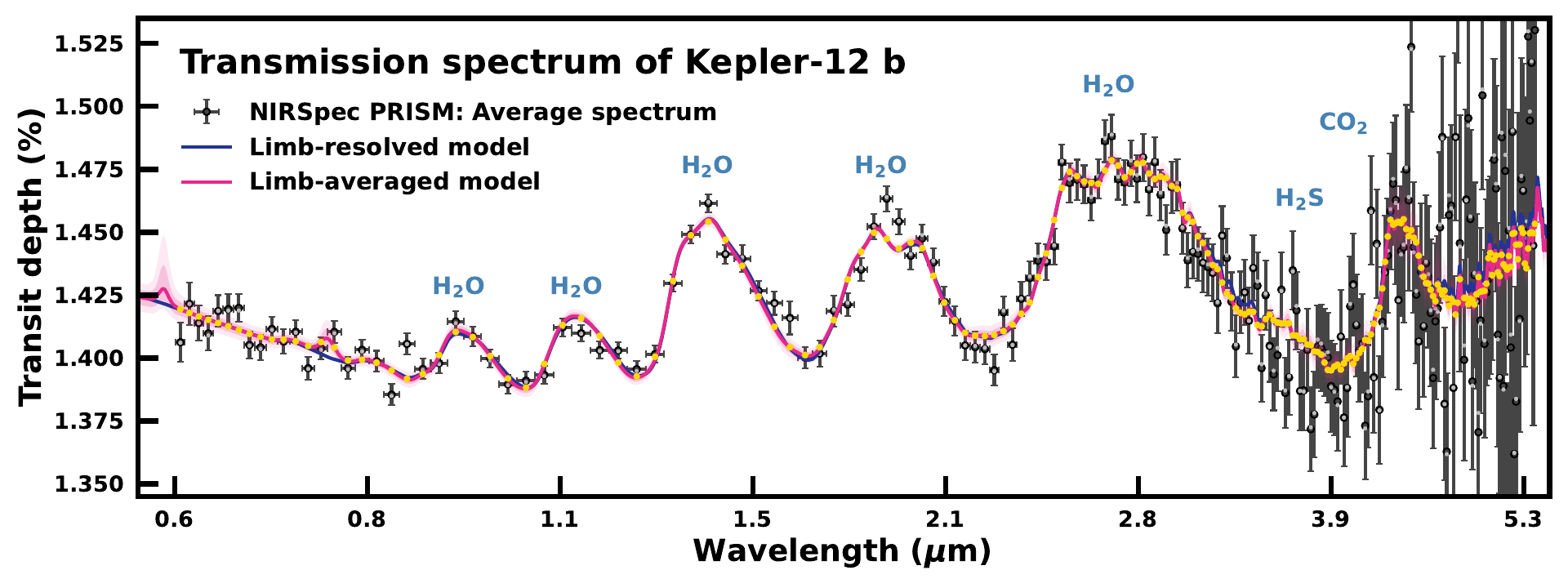}
    \vspace{-2mm}
    \caption{ The near-infrared transmission spectrum of Kepler-12~b, obtained by averaging the spectra from the two visits. The limb-averaged spectrum, obtained using \texttt{batman}, is shown by the dark grey data points, while the light grey points show the sum of the morning and evening spectra. The error bars correspond to the square root of the diagonal elements of the covariance matrix, while the full covariance matrix is used in the retrieval likelihood. The pink curve shows the median model from the normal limb-averaged retrieval, which uses the limb-averaged data shown in dark grey, while the dark blue curve shows the median of the sum of the retrieved morning and evening models from the limb-resolved retrieval. The pink shaded regions correspond to the 1--2$\sigma$ intervals of the limb-averaged model. The agreement between the two models illustrates the consistency between the limb-averaged and limb-resolved retrievals; no offset has been applied. The limb-resolved retrieval includes a reduced set of chemical species, excluding Na and K, as described in Section~\ref{sec:asymmetric}. Prominent spectral features are labelled.
    }
    \label{fig:normal_spectrum} 
\end{figure*}

After characterizing the system parameters (see Section~\ref{sec:orbital_properties}), correlated-noise properties (Appendix~\ref{app:noise}), limb darkening (Appendix~\ref{app:limb-darkening}), and exponential trends (Appendix~\ref{app:trend}), we fit the spectroscopic light curves at native spectral resolution (binning every two detector pixels). For the limb-averaged analysis, we adopt the symmetric system parameters, while for the limb-resolved analysis we adopt the asymmetric solution, as described in Section~\ref{sec:orbital_properties}. Each spectroscopic light curve is fit independently using the same framework as for the white-light analysis. In addition to the transit depths, we fit for an exponential trend amplitude and a white-noise error inflation term for each wavelength channel, resulting in 3-4 free parameters per light curve. All other parameters are fixed. We fit the light curves of the two visits separately.

A key aspect of our analysis is the treatment of covariance between the inferred transit depths (across different wavelengths, limbs, and visits). In particular, the covariance structure of limb-resolved spectra cannot be neglected without substantial information loss, as discussed in Section~\ref{sec:atmo_covariance}. We express the total covariance matrix as
\begin{equation}
\mathbf{\Sigma} = \mathbf{\Sigma}_\mathrm{common} + \mathbf{\Sigma}_\mathrm{individual},
\end{equation}
where $\mathbf{\Sigma}_\mathrm{individual}$ describes the covariance from the individual spectroscopic light-curve fits, while $\mathbf{\Sigma}_\mathrm{common}$ captures covariance induced by shared model parameters, such as orbital and limb darkening parameters. In the limb-resolved case, the covariance matrix contains strong negative correlations between the inferred morning and evening transit depths, as described in \cite{Espinoza2021}.

To estimate $\mathbf{\Sigma}_\mathrm{common}$, we propagate uncertainties in the shared model parameters into the transmission spectra. We do this by sampling from the posterior distributions of the common parameters, re-fitting the spectroscopic light curves for each realisation, and computing the sample covariance of the resulting ensemble of spectra. In our case, the common parameters are the limb-darkening parameters ([Fe/H], $\log (g)$, T$_\mathrm{eff}$, $u_{1,\,\mathrm{offset}}$, $u_{2,\,\mathrm{offset}}$; see Appendix~\ref{app:limb-darkening}),  timescale parameters for the exponential trends ($p_1$, $p_2$; see Appendix~\ref{app:trend}), and the system parameters (P, T$_0$, $b$, $\rho_\star$, $\sqrt{e} \cos \omega$, $\sqrt{e} \sin \omega$). This approach naturally maps uncertainties in the shared parameters onto the inferred transmission spectra and can be expanded to include spot-crossing parameters. In total, we generate 1000 realisations of both the limb-averaged and limb-resolved transmission spectra for each visit. These realisations are obtained using maximum likelihood estimation rather than nested sampling, as the latter did not provide sufficient precision for robust covariance estimation\footnote{The precision of the posterior-sample medians was insufficient to robustly resolve the small differences between different transmission spectra realisations.}. The final transmission spectra which we use for the atmospheric retrievals correspond to the median of the resulting ensemble of realisations, which is consistent with the standard single-fit spectra within a few ppm, well below the measurement uncertainties.

We also combine the spectra from the two visits to perform time-averaged atmospheric retrievals. For this, we compute the unweighted average of the spectra. We propagate the full covariance through the averaging using the corresponding linear transformation $\mathbf{d}' = \mathbf{A}\mathbf{d}$, giving
\begin{equation} \label{eq:cov_propagation}
\mathbf{\Sigma}' = \mathbf{A} \mathbf{\Sigma} \mathbf{A}^T\,.
\end{equation}
Here, $\mathbf{\Sigma}$ is the full covariance matrix of the data $\mathbf{d}$, including correlations across wavelengths, visits, and limbs. For the limb-averaged case, $\mathbf{d} = (\mathbf{d}^{(1)}, \mathbf{d}^{(2)})$, while for the limb-resolved case, $\mathbf{d} = (\mathbf{d}_{\mathrm{E}}^{(1)}, \mathbf{d}_{\mathrm{M}}^{(1)}, \mathbf{d}_{\mathrm{E}}^{(2)}, \mathbf{d}_{\mathrm{M}}^{(2)})$. We show the visit-combined limb-averaged transmission spectrum in Figure~\ref{fig:normal_spectrum}.

Finally, we confirm that the sum of the morning and evening spectra is consistent with the limb-averaged spectrum (as obtained using \texttt{batman}), as illustrated in Figure~\ref{fig:normal_spectrum}. The propagated covariance (after repeated use of equation~\eqref{eq:cov_propagation}) is also consistent, given that the correlation coefficients agree within 0.01 and the square roots of the diagonal elements agree to within a few per cent.

\section{Atmospheric Retrievals} \label{sec:retrievals}

In this section, we describe the atmospheric retrievals of the Kepler-12~b data presented in this work. Our retrievals are carried out in three stages: first, we analyse the limb-averaged transmission spectrum; second, we examine the limb-resolved spectra; and third, we investigate potential time variability. 

\subsection{Retrieval setup} \label{sec:retrieval_setup}
We adopt the forward-modelling framework of \texttt{POSEIDON} \citep[version 1.3.2;][]{MacDonald2017, MacDonald2022, MacDonald2024} as implemented in Appendix~\ref{app:model_implementation} and develop a custom retrieval framework to jointly fit the morning and evening spectra. Our forward models use 100 pressure layers, log-uniformly spaced between $10^{-6}$ and $10^{2}$~bar, and are generated at a spectral resolution of $R = 20,000$ with opacity sampling. We include opacity from the following gaseous species: H$_2$O, CO, CO$_2$, H$_2$S, SO$_2$, CH$_4$, NH$_3$, HCN, C$_2$H$_2$, Na, and K \citep{Polyansky2018, Li2015, Yurchenko2020, Azzam2016, Underwood2016, Yurchenko2024, Coles2019, Barber2014, Chubb2020, Ryabchikova2015}. These are selected based on expectations from thermochemical equilibrium and photochemistry in hot H$_2$-rich atmospheres \citep[e.g.,][]{Madhusudhan2016}, as well as their significant opacity in the near infrared.

For the free chemistry retrievals, we assume a constant mixing ratio with height. In terms of clouds/haze, we adopt the Mie-scattering fuzzy deck parameterisation from \cite{Zhang2019}, as implemented in \texttt{POSEIDON} \citep{Mullens2024}, parameterised by 5 parameters: the cloud top pressure of an opaque cloud ($\log P_{\mathrm{top}}$), the maximum aerosol number density ($\log n_{\mathrm{max}}$), the mean aerosol particle size ($\log (r_{\mathrm{m}}/\mu \mathrm{m})$), and the fractional scale height relative to the scale height of the gas ($f$). Here, $f$ determines how rapidly the aerosol number density $n$ falls off with height $h$, via
\begin{equation}
    \log n = \log n_{\mathrm{max}} - \frac{h}{f H_\mathrm{gas}}\,,
\end{equation}
where $H_\mathrm{gas}$ is the scale height of the background gas; such that $f = 1$ implies an approximately constant mixing ratio. We assume a log-normal particle size distribution with a width of 0.5 dex, as standard in \texttt{POSEIDON}. Given that the terminator could be inhomogeneous, we used a patchy cloud model \citep{Line2016} for the normal limb-averaged transmission spectrum, introducing one additional parameter ($\phi_\mathrm{clouds}$). Depending on the retrieval, as described below, we adopt KCl, ZnS, MnS, MgSiO$_3$, or Fe aerosols \citep{Wakeford2015, Kitzmann2018}.

For the limb-resolved retrievals, we jointly fit the morning and evening spectra using one forward model for each hemisphere\footnote{Which we multiply by $1/2$ before comparing with the data.}. Since the spectra exhibit strong covariance, we adopt a multivariate normal likelihood function \citep[similar to ][]{Espinoza2021},  defined as
\begin{equation} \label{eq:likelihood}
\begin{aligned}
    \ln L &= - \frac{1}{2} \left[ (\chi^2 +\ln  |\mathbf{\Sigma}| + k \ln (2\pi) \right] \, ,
    \end{aligned}
\end{equation}
with
\begin{equation} \label{eq:chi2}
\begin{aligned}
    \chi^2 &= (\mathbf{d} - \mathbf{m})^T \mathbf{\Sigma}^{-1} (\mathbf{d} - \mathbf{m})  \,,
    \end{aligned}
\end{equation}
where $\mathbf{d}$ and $\mathbf{m}$ denote the data and model vectors, respectively, and where $\mathbf{\Sigma}$ is the $k\times k$ covariance matrix. Here, $k$ equals the number of data points, which is doubled for the limb-resolved retrievals compared to the limb-averaged retrievals.

All retrievals are performed using the native-resolution transmission spectra. We set the planetary mass of Kepler-12 b to $M_\mathrm{p} = 137 \, M_\oplus$ \citep{Fortney2011} and fix the radius to the average white-light radius, while fitting for the reference pressure $P_\mathrm{ref}$. The complete list of parameters and priors are given in Appendix~\ref{app:retrieval_constraints} (Tables~\ref{tab:normal_atm_params}). Posterior distributions are sampled using \texttt{MultiNest} with 1000 live points.

In addition to free-chemistry retrievals, we perform retrievals under the assumption of chemical equilibrium using the limb-resolved spectra. For these models, molecular abundances are computed using \texttt{fastchem} \citep{Stock2018, Stock2022, Kitzmann2024}, without condensation, at each model evaluation. Rather than fitting individual molecular abundances, we retrieve the elemental abundances of carbon, oxygen, and sulfur relative to solar values \citep{Asplund2021}, allowing each element to vary between $10^{-2}$ and $10^{2}$ times solar. The molecular abundances are calculated on a pressure grid with a resolution of 0.5~dex, which we then interpolate onto our finer grid with 100 layers.

\subsection{Limb-averaged retrievals} \label{sec:ret:normal}

\begin{figure*}
\centering
	\includegraphics[width=1\textwidth]{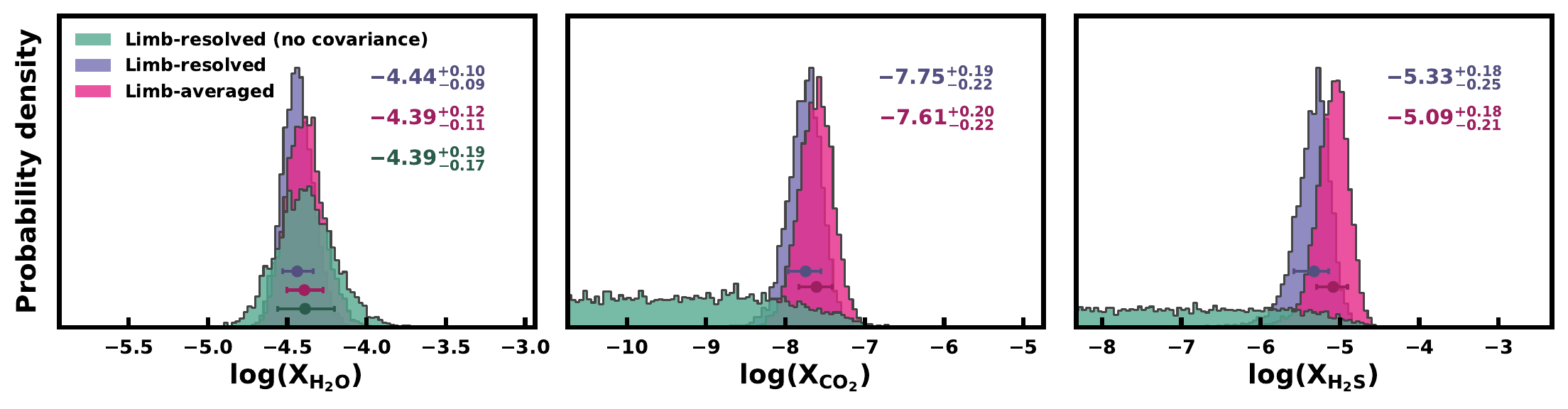}
    \vspace{-5mm}
    \caption{ Marginal posterior distributions of the log mixing ratios of H$_2$O, CO$_2$, and H$_2$S in the atmosphere of Kepler-12~b for different retrieval scenarios. The constraints obtained from the retrieval of the limb-averaged transmission spectrum are indicated in pink. The posteriors shown in purple and green represent the constraints from the limb-resolved spectrum, with and without accounting for covariance, respectively.
    }
    \label{fig:normal_abundances} 
    \vspace{1mm}
\end{figure*}

We begin with the time- and limb-averaged transmission spectrum, corresponding to the mean of the standard spectra of both visits, adopting a free-chemistry model that nominally features an isothermal pressure-temperature (P-T) profile. We conduct atmospheric retrievals using a range of aerosol species, including KCl, ZnS, MnS, MgSiO$_3$, and Fe, which may exist in the observable atmosphere based on the planet's equilibrium temperature. We also allow for patchy clouds. Our findings indicate that all aerosol species, except MnS, offer a comparable fit to the data. The abundances and temperature constraints are consistent across these aerosol species, with the exception of MnS, which is somewhat disfavoured, showing $\ln B$ values between -2.2 and -2.5 compared to the other species. Adopting MnS yields abundances and temperature constraints that differ by $\sim$2$\sigma$. When adopting KCl, ZnS, MgSiO$_3$, or Fe, we find their $\ln B$ values within 0.3 of one another, indicating that none of these species is statistically favoured over the others. Among these four species, we observe varying aerosol abundances and mean particle sizes. Overall, we establish a maximum 95\% upper limit\footnote{All upper limits mentioned in this work are provided as 95\% upper limits.} of $\log (r_{\mathrm{m}}/\mu \mathrm{m}) \,< -0.68$. Moving forward, we select MgSiO$_3$ as our canonical aerosol species. The model fit and abundance constraints are illustrated in Figures~\ref{fig:normal_spectrum} and \ref{fig:normal_abundances}, respectively.

We detect H$_2$O ($\ln B = 350.9$) and CO$_2$ ($\ln B = 8.2$) in the transmission spectrum of Kepler-12~b, together with a weaker inference of H$_2$S at $\ln B = 4.1$. For these species, we find the following abundance constraints: $\log (X_{\mathrm{H}_2\mathrm{O}}) = -4.39_{-0.11}^{+0.12}$, $\log (X_{\mathrm{CO}_2}) = -7.61_{-0.22}^{+0.20}$, and $\log (X_{\mathrm{H}_2\mathrm{S}}) = -5.09_{-0.21}^{+0.18}$, as provided in Table~\ref{tab:abundance_params}. We do not find evidence for other gas species. We retrieve an isothermal temperature constraint of $T_\mathrm{iso} = 1346_{-74}^{+69}$~K. All atmospheric constraints are presented in Table~\ref{tab:normal_atm_params}.

In addition to testing various aerosol species, we also examined several other model assumptions. Our results show little evidence for a more complex pressure-temperature (P-T) profile. Adopting the 6-parameter P-T profile from \cite{Madhusudhan2009} results in somewhat lower evidence ($\ln B = -1.4$) and produces a profile that is consistent with being isothermal. Moreover, the retrieved temperature at 1~mbar is consistent with the isothermal temperature constraint given above. We also find that the abundance constraints are fully consistent between these two P-T profiles. We also conducted a retrieval that included the effects of stellar heterogeneities \citep{Rackham2018}, but found no evidence to support this effect ($\ln B = -1.6$), which again resulted in consistent abundance constraints. For this, we adopted uniform priors for the spot temperature between $6100\pm2000$~K, a uniform prior between zero and one for the spot covering fraction, and a normal prior for the photospheric temperature of $6100\pm100$~K. Furthermore, a model that excluded aerosols provided a much poorer fit to the data ($\ln B = -46.0$), indicating strong evidence for their presence. Similarly, setting the cloud fraction to unity also led to a significantly worse fit to the data ($\ln B = -9.1$).

All retrievals were performed using the likelihood described in equation~\eqref{eq:likelihood}, which accounts for covariance arising from uncertainties in common parameters, as detailed in Sections~\ref{sec:spectrum} and \ref{sec:retrieval_setup}. Notably, setting $\mathbf{\Sigma}_\mathrm{common} = 0$ resulted in consistent constraints and did not yield significantly lower evidence ($\ln B = -0.7$). This finding suggests that the limb-averaged transmission spectrum of Kepler-12~b is not sensitive to uncertainties from common parameters, such as limb darkening or system parameters.

\begin{table*}
\renewcommand{\arraystretch}{1.65} 
\setlength{\tabcolsep}{3.3pt}
\centering
\begin{tabular}{lccccccccccc}
\hline \hline
Retrieval & H$_2$O & CO$_2$ & H$_2$S & CO & SO$_2$ & CH$_4$ & NH$_3$ & HCN & C$_2$H$_2$ & Na & K \\ \hline
Limb-averaged & $-4.39_{-0.11}^{+0.12}$ & $-7.61_{-0.22}^{+0.20}$ & $-5.09_{-0.21}^{+0.18}$ & $< -4.8$ & $< -7.1$ & $< -8.3$ & $< -6.8$ & $< -8.0$ & $< -8.1$ & $< -10.1$ & $< -8.4$ \\
Limb-resolved & $-4.44_{-0.09}^{+0.10}$ & $-7.75_{-0.22}^{+0.19}$ & $-5.33_{-0.25}^{+0.18}$ & $< -5.7$ & -- & --  & -- & --  & -- \\ \hline

\end{tabular}
\vspace{1mm}
\caption{ Retrieved abundance constraints in log volume mixing ratios. Only H$_2$O, CO$_2$, and H$_2$S are detected for which we provide the median and $\pm 1\sigma$ constraints. For other species, we report the 95\% upper limit. For the limb-resolved case presented here, we assumed symmetrical gas abundances across the morning and evening limbs using a subset of species (H$_2$O, CO$_2$, H$_2$S, and CO), as described in Section~\ref{sec:asymmetric}.
}
\vspace{-3mm}
\label{tab:abundance_params} 
\end{table*}

\subsection{Limb-resolved retrievals} \label{sec:asymmetric}

\begin{figure*}
\centering
	\includegraphics[width=1\textwidth]{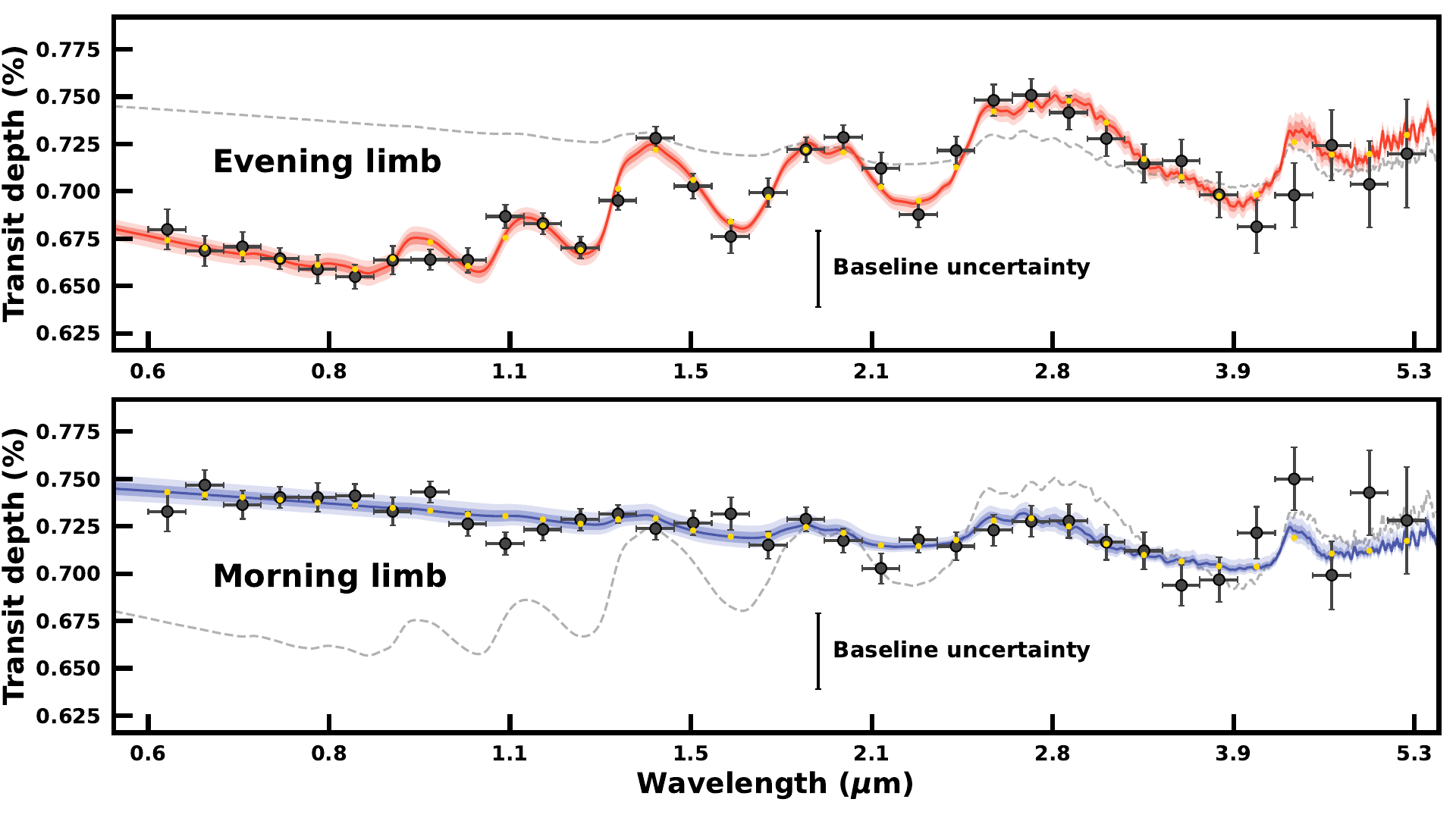}
    \vspace{-5mm}
    \caption{Limb-resolved transmission spectra of Kepler-12~b, binned to $R = 15$ for clarity. The plotted uncertainties correspond to the (binned) diagonal elements of $\mathbf{\Sigma}_\mathrm{individual}$ (see Section~\ref{sec:spectrum}). The blue and red curves show the median free-chemistry model for the morning and evening limbs, respectively, assuming an isothermal P-T profile and equal abundances between the two limbs. The shaded regions correspond to the model's 1--2$\sigma$ intervals. The yellow points represent the binned model. The uncertainties in the overall offsets between the two spectra, derived from $\mathbf{\Sigma}_\mathrm{common}$, which we refer to as baseline uncertainties, are illustrated by the black error bars (201 ppm and strongly anti-correlated). Due to these baseline uncertainties, the model can be offset from the data while still providing a good fit. For visual comparison, we shift the morning and evening model spectra by $+67$~ppm and $-67$~ppm, respectively. The retrievals were performed on the unshifted spectra, and these shifts are applied only for visualization. The offsets were chosen to approximately align the model and data and are small compared with the $\sim$200~ppm baseline uncertainties. Grey dashed curves show the median morning and evening model spectra, to facilitate direct comparison.}
    \label{fig:asymmetric_spectrum} 
\end{figure*}

\begin{figure*}
\centering
	\includegraphics[width=1\textwidth]{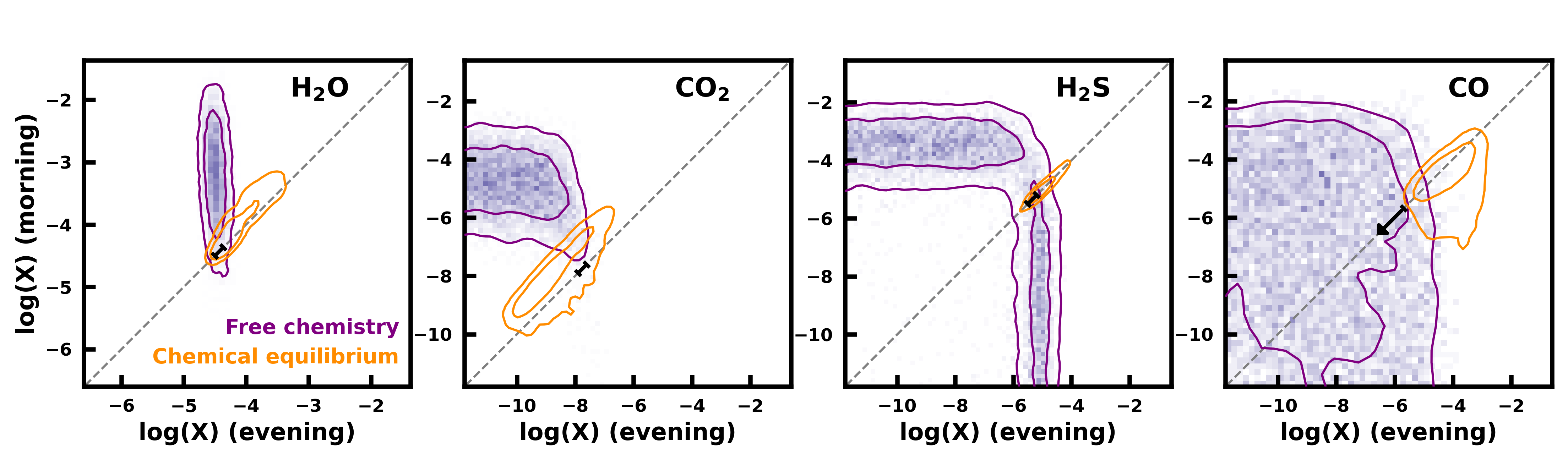}
    \vspace{-4mm}
    \caption{ Two-dimensional posterior distributions of the mixing ratios of H$_2$O, CO$_2$, H$_2$S, and CO, showing the constraints for each limb. The purple and orange contours represent the 68\% and 95\% probability regions for the free chemistry and chemical equilibrium retrievals, respectively, where the chemical equilibrium case assumes equal elemental abundances between the limbs. The purple regions indicate the posterior distribution for the free chemistry retrieval. The chemical equilibrium abundances are presented at a pressure of 1 mbar, which approximately corresponds to the pressure probed through transmission spectroscopy. The diagonal dashed lines shows the lines of equal mixing ratios between the morning and evening limbs. The black error bars for H$_2$O, CO$_2$, and H$_2$S show the free chemistry constraints on these species under the assumption of equal abundances between the limbs. For CO, we show the 95\% upper limit.
    }
    \label{fig:2d_abundances} 
\end{figure*}

We jointly fit the average morning and evening spectra of Kepler-12~b while accounting for the non-negligible covariance between data points. The covariance matrix is presented in Appendix~\ref{app:cov}. We adopt a model including H$_2$O, CO$_2$, CO, and H$_2$S in the gas phase; and MgSiO$_3$ aerosols. We select this subset of gaseous species given that these are either detected in the normal spectrum (H$_2$O, CO$_2$, H$_2$S) or are expected to be abundant in the atmosphere (CO). Adopting KCl, ZnS, or Fe aerosols does not significantly change the retrieved abundances or temperature constraints, as discussed below. We present the morning and evening spectra of Kepler-12~b in Figure~\ref{fig:asymmetric_spectrum}. 

To begin, we conduct a free-chemistry retrieval where we allow all atmospheric properties to vary between the two limbs, except the reference pressure $P_\mathrm{ref}$, which we keep as a common parameter\footnote{Allowing $P_\mathrm{ref}$ to vary between the limbs produce consistent results for the retrieved $P_\mathrm{ref}$ values along with all other atmospheric parameters. Similarly, setting $P_\mathrm{ref}=10$~bar and fitting for a common R$_\mathrm{p}$ \citep[e.g.,][]{Welbanks2022} also give consistent atmospheric constraints.}. In total, this model has 19 free parameters. We evaluate the significance of potential chemical asymmetries by performing a series of retrievals in which we fix the gas abundances to be identical between the two limbs, one species at a time. Generally, we do not find significant evidence for abundance asymmetries between the morning and evening limbs ($\ln B < 1.3$, where the maximum occurs for CO$_2$). However, the morning limb shows weak preference for higher CO$_2$ abundances. The two-dimensional marginal posterior distributions of the log mixing ratios are presented in Figure~\ref{fig:2d_abundances}. Although the marginal posteriors for some species peak away from the diagonal, which represents equal abundances, the Bayesian evidence provides only weak support for allowing independent abundances between the limbs. For H$_2$O, we obtain $\log (X_{\mathrm{H}_2\mathrm{O},\,\mathrm{E}}) = -4.49_{-0.10}^{+0.10}$ and $\log (X_{\mathrm{H}_2\mathrm{O},\,\mathrm{M}}) = -3.20_{-0.69}^{+0.64}$. For CO$_2$, the abundance on the evening limb has a 95\% upper limit of $\log (X_{\mathrm{CO}_2,\,\mathrm{E}}) < -8.1$, while on the morning limb, we find $\log (X_{\mathrm{CO}_2,\,\mathrm{M}}) = -4.89_{-0.97}^{+0.91}$. Despite this apparent contrast, the full joint posterior retains a tail towards equal CO$_2$ abundances, with approximately 99.3\% of the posterior favouring a higher abundance on the morning limb. This corresponds to a significance of approximately $2.7\sigma$ (two-sided), while the Bayesian evidence ($\ln B = 1.3$) indicates that the improvement of the fit does not strongly justify the additional freedom of independent limb abundances.

H$_2$S has a posterior distribution that allows for its presence on either the morning or evening limbs, or both together. The anti-correlation between the two limb spectra allows the H$_2$S absorption to be distributed between the limbs while producing a similar combined spectrum. However, removing H$_2$S from both limbs simultaneously cannot reproduce the observed feature in the combined spectrum. We obtain 95\% upper limits of $\log (X_{\mathrm{H}_2\mathrm{S},\,\mathrm{E}}) < -4.9$ and $\log (X_{\mathrm{H}_2\mathrm{S},\,\mathrm{M}}) < -2.7$.

Finally, we find that a model which shares the abundances of all gaseous species between the two limbs produces a comparable fit to the data and yields abundance constraints that are consistent with the analysis of the limb-resolved spectrum, as shown in Table~\ref{tab:abundance_params} and Figure~\ref{fig:normal_abundances}. Therefore, given the weak Bayesian evidence for abundance asymmetries, we use symmetric abundances going forward unless otherwise stated. Additional observations are required to robustly determine whether the atmosphere exhibits inhomogeneous abundances.

\subsubsection{Aerosol properties}

We observe a significant difference in aerosol properties between the morning and evening limbs. A model that excludes aerosols on the morning limb is strongly disfavoured ($\ln B = -44.0$), whereas a model without aerosols on the evening limb is weakly preferred ($\ln B = 1.4$). We therefore find no evidence for aerosols in the observable atmosphere of the evening terminator, while aerosols are required to explain the morning spectrum. This result is consistent with the strong preference for patchy aerosols in the limb-averaged analysis presented in Section~\ref{sec:ret:normal}. Although, unlike the limb-averaged analysis, adopting patchy clouds here yields somewhat lower evidence ($\ln B = -2.1$) than the nominal aerosol model and does not significantly affect the retrieved abundances or temperature structure. The  morning-limb cloud fraction is constrained to be $\phi_\mathrm{M} = 0.884_{-0.107}^{+0.076}$, while the evening-limb cloud fraction remains largely unconstrained. The total cloud fraction, retrieved to be $\phi = (\phi_\mathrm{M} + \phi_\mathrm{E})/\,2 = 0.414_{-0.043}^{+0.045}$ using the limb-averaged spectrum (see Table~\ref{tab:normal_atm_params}), is consistent with $\phi_\mathrm{M}/2 = 0.442_{-0.054}^{+0.038}$, assuming $\phi_\mathrm{E} = 0$ in the observable atmosphere. 

We further consider a retrieval including MgSiO$_3$, KCl, ZnS, and Fe aerosols on both limbs. This model yield consistent abundance and temperature constraints and weakly favour KCl as the dominant aerosol species. However, this preference is not statistically significant, as replacing MgSiO$_3$ with KCl alone results in only a marginal improvement in fit quality ($\ln B = 0.5$). We therefore adopt MgSiO$_3$ as our fiducial aerosol species, as it is likely more physically plausible under the atmospheric conditions of Kepler-12 b; KCl condensation generally requires substantially lower temperatures. Importantly, the choice of aerosol species does not affect our conclusions regarding the atmospheric temperature structure or molecular abundances.

For the morning limb, the MgSiO$_3$ retrieval yields a mean particle size of $\log (r_{\mathrm{m,\,M}}/\mu \mathrm{m}) = -1.23_{-0.27}^{+0.22}$, while the KCl retrieval provides only an upper limit of $\log (r_{\mathrm{m,\,M}}/\mu \mathrm{m}) < -1.03$. The MgSiO$_3$ constraint is consistent with the limb-averaged result of $\log (r_{\mathrm{m}}/\mu \mathrm{m}) = -0.98_{-0.33}^{+0.21}$ (which also adopted MgSiO$_3$). Opaque cloud decks are not invoked to explain the data, with 95\% limits of $\log (\mathrm{P}_\mathrm{deck,\,E}/\mathrm{bar}) > -0.65$ and $\log (\mathrm{P}_\mathrm{deck,\,M}/\mathrm{bar}) > -1.0$. Overall, we find strong evidence for aerosols on the morning limb and retrieve consistent aerosol properties between the limb-averaged and limb-resolved analysis, but we are unable to definitively identify the aerosol species responsible. Upcoming observations using MIRI LRS (JWST GO 7849), may be able to better constrain which aerosol species are present.

\subsubsection{Temperature structure} \label{sec:temperature}

\begin{figure}
\centering
	\includegraphics[width=0.49\textwidth]{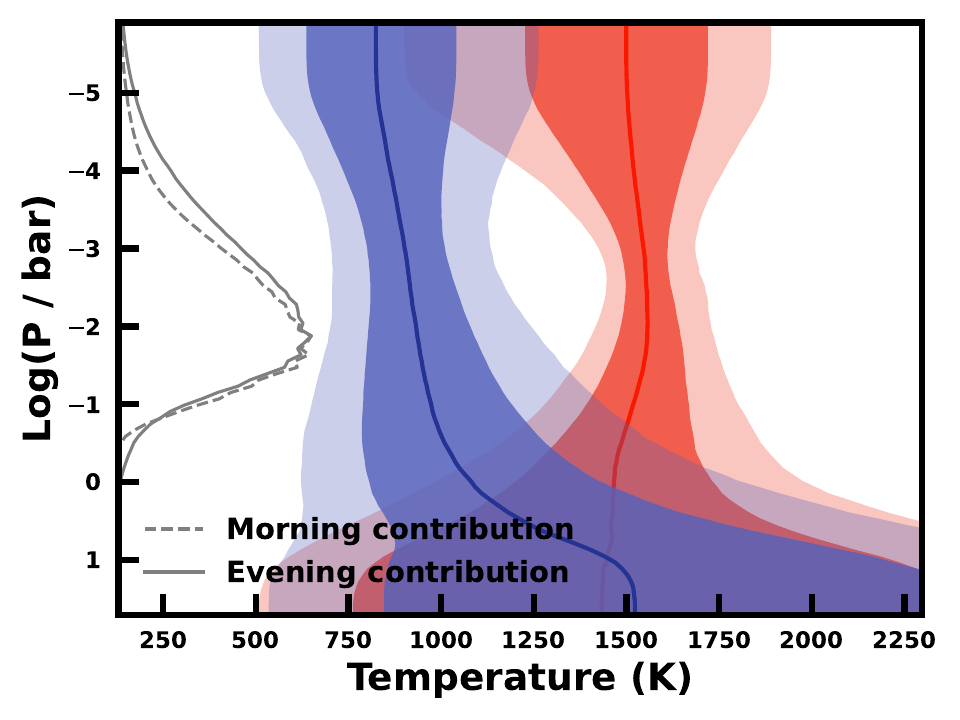}
    \vspace{-5mm}
    \caption{ Retrieved P-T profiles for the morning and evening spectra, shown in blue and red, respectively. These profiles were obtained by adopting the three-point P-T profile described in Section~\ref{sec:temperature}. The coloured lines correspond to the median model, while the shaded regions represent the 1–2$\sigma$ intervals. The grey dashed and solid lines represent the median contribution function for the morning and evening spectra, respectively.
    }
    \label{fig:P-T_profile} 
\end{figure}

A significant temperature gradient is observed between the morning and evening limbs. For the isothermal model in which the chemical abundances are shared between the two limbs, we retrieve an evening temperature of $\mathrm{T}_\mathrm{iso,\,E} = 1545_{-54}^{+54}$~K and a morning temperature of $\mathrm{T}_\mathrm{iso,\,M} = 881_{-94}^{+100}$~K, corresponding to a temperature difference of $\Delta \mathrm{T}_\mathrm{iso} = 660_{-110}^{+110}$~K.

To assess the impact of the assumed P-T profile parameterisation, we also consider non-isothermal atmospheres. We first adopt a two-point profile defined by temperatures at fixed pressures of $P_\mathrm{deep} = 10$~bar and $P_\mathrm{high} = 10^{-5}$~bar, with the temperature varying linearly in $\log P$ between these points and remaining isothermal outside this region. We then consider a three-point profile by introducing an additional temperature node at a free pressure, $P_\mathrm{mid}$.  These profiles were then smoothed to remove sharp edges, as implemented in \texttt{POSEIDON}. For these models, we adopt the priors $T_\mathrm{deep} \sim \mathcal{U}(450~\mathrm{K}, 3000~\mathrm{K})$, $T_\mathrm{mid} \sim \mathrm{U}(450~\mathrm{K}, 2000~\mathrm{K})$, $T_\mathrm{high} \sim \mathcal{U}(450~\mathrm{K}, 2000~\mathrm{K})$, and $\log(P_\mathrm{mid}/\mathrm{bar}) \sim \mathcal{U}(-5,1)$. The pressure $P_\mathrm{mid}$ is shared between the two limbs. The retrievals assume a common $P_\mathrm{ref}$ between the two limbs; allowing $P_\mathrm{ref}$ to vary between the limbs yields consistent thermal profiles.

The non-isothermal models are weakly disfavoured relative to the isothermal model, with $\ln B = -1.2$ and $\ln B = -2.5$ for the two-point and three-point profiles, respectively. Nevertheless, the retrieved molecular abundances and aerosol properties remain fully consistent across all P-T parameterisations. The temperatures near the photosphere, at 1~mbar, are similarly robust: for the two-point profile we obtain T$_{\mathrm{1\,mbar},\,\mathrm{E}}=1541_{-63}^{+68}$~K and T$_{\mathrm{1\,mbar},\,\mathrm{M}}=910_{-110}^{+110}$~K, while for the three-point profile we obtain T$_{\mathrm{1\,mbar},\,\mathrm{E}}=1546_{-63}^{+65}$~K and T$_{\mathrm{1\,mbar},\,\mathrm{M}}=900_{-100}^{+110}$~K. These correspond to temperature differences of $\Delta \mathrm{T}_\mathrm{1\,mbar} = 640_{-140}^{+130}$~K and $\Delta \mathrm{T}_\mathrm{1\,mbar} = 650_{-140}^{+130}$~K, respectively. We therefore conclude that the inferred temperature contrast at the photosphere is insensitive to the assumed P-T profile complexity. We adopt the retrieved P-T profile from the three-point model as our fiducial temperature constraints, shown in Figure~\ref{fig:P-T_profile}.

Allowing the chemical abundances to vary independently between the two limbs leads to a somewhat lower temperature on the morning side, although the results remain consistent within $2\sigma$. In this case, we retrieve $\mathrm{T}_\mathrm{iso,\,E} = 1582_{-61}^{+60}$~K and $\mathrm{T}_\mathrm{iso,\,M} = 696_{-76}^{+92}$~K, corresponding to a larger temperature contrast of $\Delta \mathrm{T}_\mathrm{iso} = 935_{-88}^{+81}$~K. Regardless of the abundance parameterisation adopted, the evening terminator is found to be significantly hotter than the morning terminator. A conservative estimate from the three-point P--T profile places the 95\% lower limit on the 1~mbar temperature difference at 408~K.

\subsubsection{Stellar heterogeneities}

As for the limb-averaged analysis, we find that stellar heterogeneities are disfavoured by the limb-resolved spectra ($\ln B = -3.2$). For this analysis, we note that unocculted stellar heterogeneities modify the morning and evening transmission spectra equally, assuming that heterogeneities does not change significantly during transit, for example, due to stellar rotation. We justify this assumption by the lack of periodic modulations in the Kepler light curves of Kepler-12 \citep{Fortney2011}.

Despite the lower evidence of this model, we recover consistent abundance constraints and aerosol properties when compared to models without stellar heterogeneities. In this case, we retrieve $\mathrm{T}_\mathrm{iso,\,E} = 1481_{-60}^{+64}$~K and $\mathrm{T}_\mathrm{iso,\,M} = 926_{-88}^{+95}$~K, corresponding to a temperature contrast of $\Delta \mathrm{T}_\mathrm{iso} = 550_{-120}^{+120}$~K. Although this is somewhat lower than what we find without stellar heterogeneities, it remains consistent within the uncertainties.

\subsubsection{Constraints on low-temperature species}

Due to the low inferred temperature on the morning limb, we run an additional retrieval including CH$_4$ and NH$_3$, where we let the abundances of all species vary between the two limbs. However, we find no evidence for the inclusion of these species. In fact, this retrieval results in lower evidence compared to the retrieval with only H$_2$O, CO$_2$, CO, and H$_2$S ($\ln B = -3.8$). For CH$_4$, we obtained a 95\% upper limit of $\log (X_{\mathrm{CH}_4,\,\mathrm{M}}) < -7.5$ on the morning side and $\log (X_{\mathrm{CH}_4,\,\mathrm{E}}) < -8.2$ on the evening side. Similarly, for NH$_3$, we found a 95\% upper limit of $\log (X_{\mathrm{NH}_3,\,\mathrm{M}}) < -5.8$ on the morning side and $\log (X_{\mathrm{NH}_3,\,\mathrm{E}}) < -6.7$ on the evening side. These results are consistent with the analysis of the normal limb-averaged transmission spectrum, where we also did not detect CH$_4$ and NH$_3$.

\subsubsection{The effects of covariance}

Finally, we evaluate the importance of accounting for the covariance by conducting a retrieval that only uses the diagonal elements of $\mathbf{\Sigma}_\mathrm{individual}$. We find that neglecting the covariance significantly reduces the sensitivity of the retrieval. In this scenario, H$_2$O is the only species for which we obtain meaningful constraints, specifically $\log(X_{\mathrm{H}_2\mathrm{O}}) = -4.39_{-0.17}^{+0.19}$. The precision on the H$_2$O log mixing ratio thus decreases by a factor of two when neglecting the covariance, while CO$_2$ and H$_2$S are not detected at all. Likewise, the precision of the morning/evening temperature difference is also reduced, with $\Delta \mathrm{T}_\mathrm{iso} = 650_{-140}^{+190}$~K. We compare these abundance constraints to the standard constraints in Figure~\ref{fig:normal_abundances}. Additionally, the evidence is considerably diminished compared to the model fit that correctly incorporates the covariance, with $\ln B = -176$. We therefore emphasise the importance of considering covariance when interpreting limb-resolved transmission spectra.

\subsection{Chemical-equilibrium retrieval} \label{sec:chem_eq}

We perform additional retrievals of the limb-asymmetry spectra, imposing chemical equilibrium on the species H$_2$O, CO$_2$, CO, and H$_2$S. These species are expected to be the dominant carbon, oxygen, and sulfur-bearing species in chemical equilibrium, and they exhibit significant spectral features in the near infrared. We do not consider other species in this analysis, such as CH$_4$, Na or K, as we obtain strict upper limits on their abundances. In total, we fit the elemental abundances of C, O, and S using log-uniform priors ranging between -2 and 2 dex relative to solar abundances. The abundances of other elements, except for H and He, are fixed to scale with oxygen. We compute the abundances separately for each limb due to the differing temperatures.

The retrieved elemental abundances are sub-solar, with $[\mathrm{C}/\mathrm{H}] = -0.98_{-0.53}^{+0.66}$, $[\mathrm{O}/\mathrm{H}] = -0.87_{-0.26}^{+0.55}$, and $[\mathrm{S}/\mathrm{H}] = -0.27_{-0.25}^{+0.40}$. These correspond to a $\mathrm{C}/\mathrm{O}$ ratio of $0.45_{-0.21}^{+0.16}$, consistent with sub-solar to solar values. We caution, however, that these inferences may be biased by disequilibrium chemistry or other unmodeled processes. The inferred mixing ratios at 1~mbar, approximately the pressure level probed, agree with the free-chemistry constraints within the 95\% probability contours (Figure~\ref{fig:2d_abundances}). The posterior distributions nevertheless differ in shape because the chemical-equilibrium retrieval imposes more restrictive assumptions, with the morning and evening abundance constraints consequently showing closer agreement.

\subsection{Time-resolved retrievals} \label{sec:time-resolved}

\begin{figure*}
\centering
	\includegraphics[width=1\textwidth]{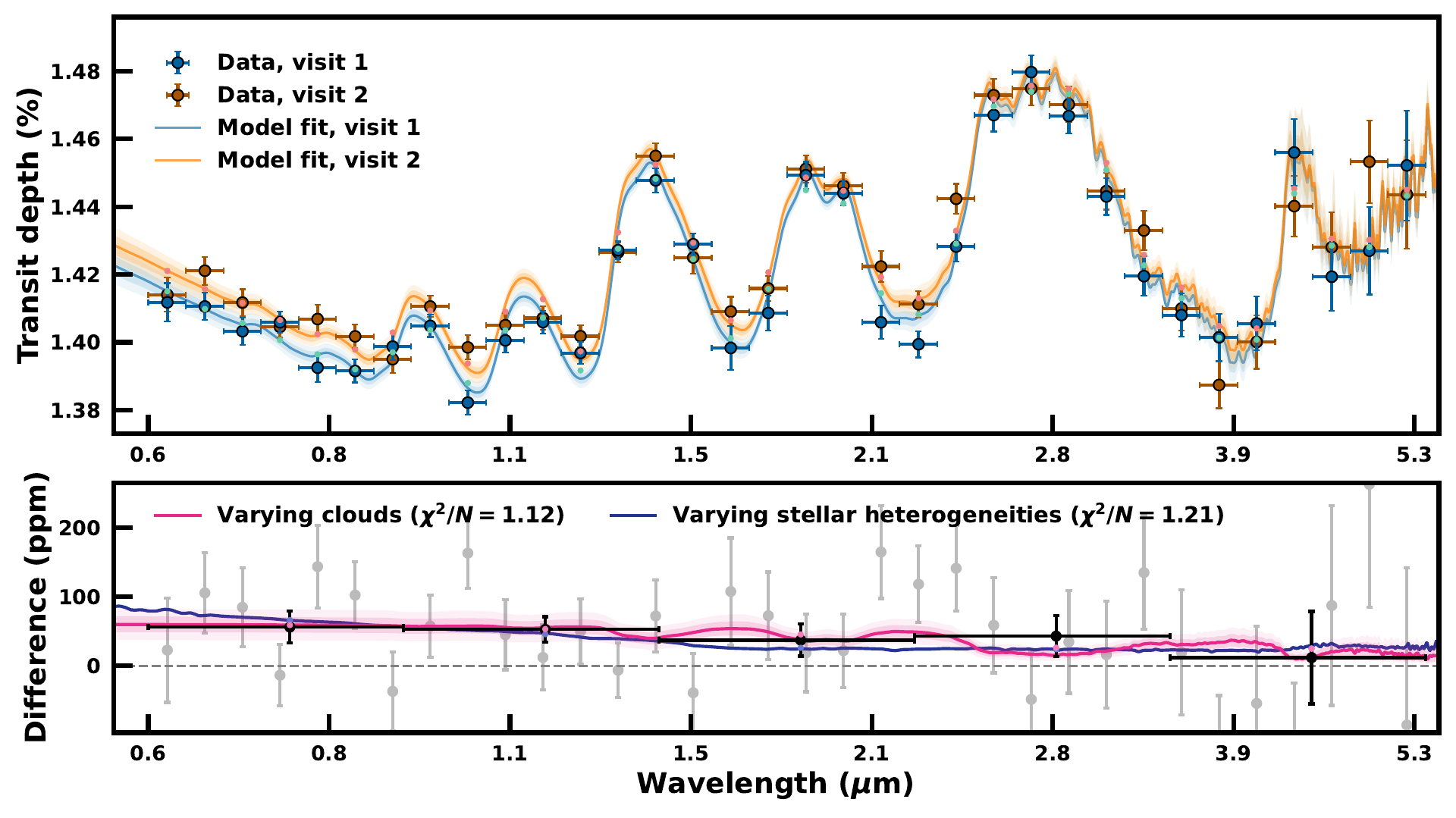}
    \vspace{-5mm}
    \caption{ Normal limb-averaged transmission spectra of Kepler-12~b from both visits. The blue and orange data points  correspond to visits 1 and 2, respectively, and are binned {to $R=15$} for visual clarity. The blue and orange curves, along with the shaded regions, indicate the median model fits and their associated 1--2$\sigma$ regions for visits 1 and 2, respectively. The model shown in the top panel is jointly fit to both spectra, where the only varying parameter is the aerosol abundance. The turquoise and pink points represent the binned model for visits 1 and 2, respectively. The bottom panel illustrates the difference between the two visits, with {$R=2$ in black and {$R=15$} in light grey. The model in pink corresponds to the varying cloud-abundance model in the top panel, while the dark blue model shows the varying stellar-heterogeneity model. We report the $\chi^2/N$ for the visit-to-visit difference, calculated using the $R=15$ data, where $N$ is the number of data points.}
    }
    \label{fig:normal_spectrum_time} 
\end{figure*}

Next, we assess the possibility of temporal variability between the two visits. As a first step, we repeat the retrievals in Sections~\ref{sec:ret:normal} and~\ref{sec:asymmetric} for each visit separately. The retrieved atmospheric constraints are provided in Appendix~\ref{app:retrieval_constraints} (see Tables~\ref{tab:normal_atm_params} and~\ref{tab:asymmetric_atm_params}). We find that the H$_2$O and CO$_2$ abundances are consistent within $\sim$$2\sigma$. H$_2$S is detected in the first visit but not in the second. Other parameters are consistent within $\sim1-2\sigma$. Moreover, as shown in Figure~\ref{fig:normal_spectrum_time} and indicated by the white-light curve analysis in Section~\ref{sec:white_time}, the transit depths differ systematically between the two visits, particularly at shorter wavelengths. Thus, although there is some tension between the retrieved atmospheric parameters, it is not immediately clear what drives the observed spectral differences or how significant these differences are. To investigate this further, we conduct a series of joint free-chemistry retrievals of the limb-averaged transmission spectra of both visits, systematically varying which parameters are allowed to change between visits.

\begin{table}
\vspace{3mm}
\renewcommand{\arraystretch}{1.55} 
\setlength{\tabcolsep}{10pt}
\centering
\begin{tabular}{lc}
\hline \hline
Retrieval Scenario & $\ln B$ \\ \hline
Varying stellar heterogeneities & 2.2 (3.3)\\
Varying gas abundances & -4.0 \\
Varying temperature & 0.1 \\
Varying aerosol abundance ($\log n_{\mathrm{max}}$) & 5.7 \\
Varying cloud coverage fraction & 6.5 \\ \hline
\end{tabular}
\vspace{1mm}
\caption{Log Bayes factors for different model assumptions when jointly fitting the transmission spectra of both visits, compared to a model with a static atmosphere. For the case involving varying stellar heterogeneities, we also computed the Bayes factor in comparison to a model that included static stellar heterogeneities, which is presented in parentheses. More details are provided in Section~\ref{sec:time-resolved}.
}
\label{tab:time_variability_normal} 
\end{table}

We restrict this analysis to the limb-averaged spectra because the inferred variability of the individual limb-resolved spectra is more sensitive to the adopted orbital assumptions, as discussed in Section~\ref{sec:white_time}. For completeness, we present the limb-resolved spectra of the two visits in Appendix~\ref{app:individual-visit} (Figure~\ref{fig:asymmetric_spectrum_time}). This approach enables us to assess the significance of the observed spectral differences and determine which properties, if any, need to vary to explain the observations, while also accounting for the covariance between the two data sets.

We perform the joint retrievals using the same subset of gaseous species as in Section~\ref{sec:asymmetric} (i.e., H$_2$O, CO$_2$, CO, and H$_2$S). First, we run a reference retrieval assuming that all parameters are constant in time. We confirm that this yields atmospheric constraints consistent with those obtained from the retrieval of the average spectrum. Next, we run a retrieval with stellar heterogeneities in which we keep the atmospheric parameters constant while allowing the spot/facula temperature and covering fraction to vary between visits. This allows us to test whether the apparent variability can be explained by changes in unocculted stellar heterogeneities between the visits. We then perform a series of tests allowing each of the following atmospheric properties to vary independently between visits: (1) the molecular abundances, (2) the isothermal temperature, (3) the cloud fraction, and (4) the maximum aerosol number density, while keeping the other parameters constant in time.

The Bayes factors for the different cases are presented in Table~\ref{tab:time_variability_normal}. When compared to the steady-state reference model, we find a strong preference for allowing either the cloud fraction ($\ln B = 6.5$) or the maximum aerosol number density ($\ln B = 5.7$) to vary between visits. Varying the molecular abundances or the temperature does not yield a better fit to the data. We also find that allowing for temporal changes in unocculted stellar heterogeneities yields a more modest increase in evidence ($\ln B = 2.2-3.3$), depending on whether the model is compared to a static model with or without stellar heterogeneities. Despite the significantly lower evidence, this model is able to reasonably fit the spectra as shown in Figure~\ref{fig:normal_spectrum_time}; suggesting that the model evidence may be penalised by the additional parameters describing the stellar heterogeneities. Overall, we find strong evidence that the transmission spectra differ between the two visits, with temporal changes in aerosol properties providing the most strongly favoured explanation among the models considered. Additional observations will be crucial to establish the physical origin of the variability.

Although the limb-resolved analysis suggests that most of the variability originates from the evening limb, this interpretation is sensitive to the adopted orbital assumptions, as discussed in Section~\ref{sec:white_time}. If the observed variability is driven by clouds, an evening-limb origin would be somewhat unexpected, given that we observe a relatively cloud-free evening limb and a cloudier morning limb. We might therefore expect cloud variability to have a larger effect on the morning limb. However, other forms of atmospheric variability not explored in our retrievals could also contribute to the observed differences. Alternatively, if the variability is stellar in origin, it could reflect subtle changes in occulted stellar heterogeneities near the stellar limb which mimic temporal variations in the inferred limb asymmetry.

\section{Summary and discussion} \label{sec:discussion}

We present the first transmission spectrum of the hot Jupiter Kepler-12~b, obtained using two JWST/NIRSpec PRISM (0.6--5.4~$\mu$m) transit observations separated by approximately two months. We detect H$_2$O ($\ln B = 350.9$) and CO$_2$ ($\ln B = 8.2$) in the planet's atmosphere and find strong evidence for pronounced limb asymmetries. We also infer the presence of H$_2$S at a lower significance of $\ln B = 4.1$. Through a joint analysis of the limb-resolved spectra, we infer a cloudy morning limb that is significantly cooler than the evening limb, which appears largely cloud-free within the observable atmosphere. While the retrieved gas-phase abundances are broadly consistent between the two limbs, the morning limb weakly favours enhanced abundances, particularly for CO$_2$. Under the assumption of chemical equilibrium and common elemental abundances across the terminator, we infer subsolar atmospheric abundances of carbon, oxygen, and sulfur, with $[\mathrm{C}/\mathrm{H}] = -0.98_{-0.53}^{+0.66}$, $[\mathrm{O}/\mathrm{H}] = -0.87_{-0.26}^{+0.55}$, and $[\mathrm{S}/\mathrm{H}] = -0.27_{-0.25}^{+0.40}$, corresponding to a C/O ratio of $0.45_{-0.21}^{+0.16}$. In addition to the spatial variations across the terminator, we find strong evidence for temporal variability in the transmission spectrum between the two observing epochs ($\ln B = 5.7-6.5$). Our retrievals suggest that the observed variability can be explained by changing cloud properties; however, variable stellar heterogeneities also fit the data and cannot be ruled out. 

\subsection{A large temperature contrast across the terminator}

\begin{figure}
\centering
	\includegraphics[width=0.49\textwidth]{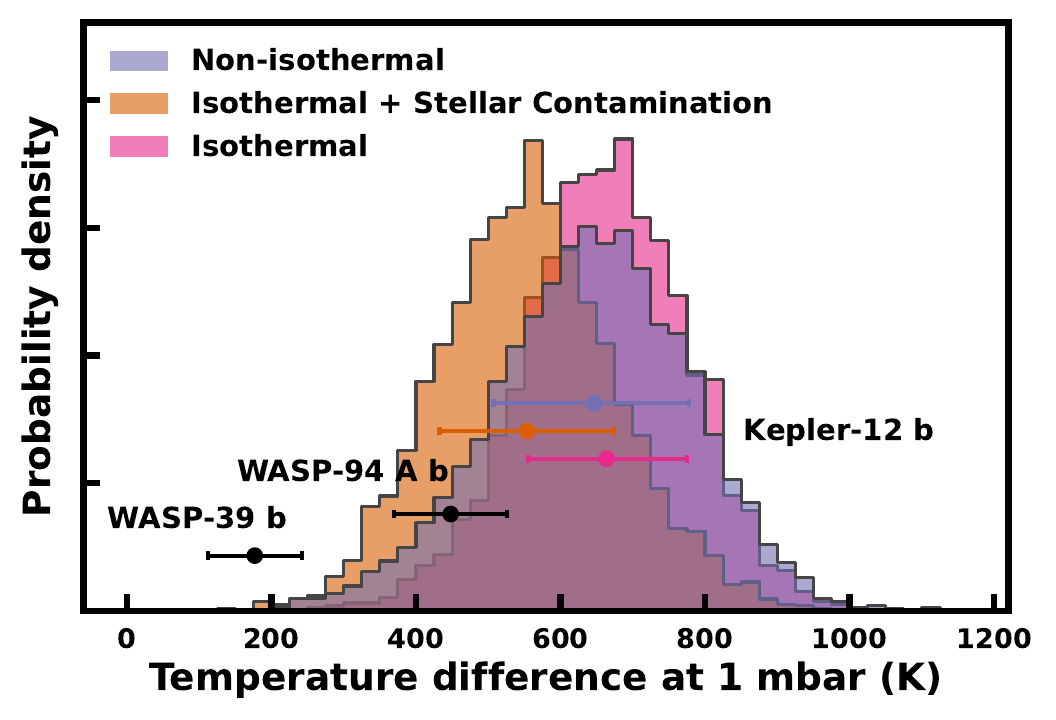}
    \vspace{-5mm}
    \caption{ Retrieved temperature difference between the morning and evening limbs of Kepler-12~b at 1~mbar, as described in Section~\ref{sec:asymmetric}.  The orange and pink posteriors come from retrievals adopting an isothermal P-T profile with and without stellar heterogeneities, respectively. Note that stellar heterogeneities is not favoured by the data but is included here for completeness. The purple posterior comes from the retrieval case using the three-point P-T profile. The black data points represent constraints for WASP-39~b \citep{Espinoza2024} and WASP-94~A~b \citep{Mukherjee2026} for comparison.
    }
    \label{fig:temp_comparison} 
\end{figure}

We retrieve a large temperature contrast between the two limbs, with $\mathrm{T}_{\mathrm{1\,mbar},\,\mathrm{E}} = 1546_{-63}^{+65}$~K and $\mathrm{T}_{\mathrm{1\,mbar},\,\mathrm{M}} = 900_{-100}^{+110}$~K (see Section~\ref{sec:temperature}). This corresponds to a difference of $\Delta \mathrm{T}_{\mathrm{1\,mbar}} = 650_{-140}^{+130}$~K, with a 95\% lower limit of $>$$408$~K. This contrast exceeds predictions from general circulation models, which typically find $\sim200$–400~K for planets with T$_\mathrm{eq} \sim 1500$~K \citep{Kataria2016, Triantafillides2025}. The inferred temperature difference is also larger than other observational constraints to date, as illustrated in Figure~\ref{fig:temp_comparison}. The closest case is WASP-94~b, with $\Delta \mathrm{T} = 448_{-75}^{+80}$~K from NIRISS SOSS observations \citep{Mukherjee2026}. Interestingly, WASP-94~b and Kepler-12~b share similar bulk properties \citep{Bonomo2017, Neveu-VanMalle2014}, including mass ($0.46$~M$_\mathrm{J}$), radius ($1.72$~R$_\mathrm{J}$), and equilibrium temperature (T$_\mathrm{eq} \approx 1600$~K), which may point to common atmospheric processes driving large terminator asymmetries.

In contrast, cooler planets exhibit smaller temperature contrasts. For example, WASP-39~b and WASP-107~b show $\Delta \mathrm{T} = 177_{-57}^{+65}$~K \citep{Espinoza2024} and $\sim$$180$~K \citep{Murphy2024b, Murphy2025}, respectively. At higher irradiation, MIRI LRS eclipse mapping of WASP-17~b (T$_\mathrm{eq} = 1770$~K) yields $\Delta \mathrm{T} \sim 500$~K \citep{Valentine2024}. Finally, \citet{Fu2025} predict $\Delta \mathrm{T} \approx 700$~K for Kepler-12~b based on trends between forward models and the observed 1.4~$\mu$m H$_2$O feature in several NIRISS SOSS spectra. Overall, these comparisons place Kepler-12~b among the most extreme cases of terminator temperature asymmetry currently known. While our conservative lower limit ($>$$408$~K) overlaps with the highest previously reported values, the nominal contrast of $\sim650$~K suggests that Kepler-12~b may represent the upper end of the distribution.

\subsection{Cloud distribution and properties}

Our retrievals reveal a strongly asymmetric aerosol distribution across the terminator. The morning limb is consistently found to be aerosol-rich in both visits, whereas the observable atmosphere of the evening limb is consistent with being clear, with no evidence supporting the inclusion of aerosols. This behaviour is qualitatively similar to WASP-94~A~b \citep{Mukherjee2026}, although in that case aerosols are inferred on both limbs, whereas for Kepler-12~b we find no evidence for aerosols on the evening limb. We also find no statistically significant preference for a specific condensate species, although MnS is disfavoured.

The retrieved cloud properties suggest a vertically extended aerosol layer composed of small particles. We place a conservative 95\% upper limit on the mean particle size of $\log(r_{\mathrm{m,,M}}/\mu\mathrm{m}) \lesssim -1.0$, consistent with the small particles predicted by cloud microphysics models \citep[e.g.,][]{Powell2019}. Such particles naturally produce the strong spectral slope observed on the morning limb while exhibiting long settling timescales that favour vertically extended cloud structures. This interpretation is further supported by the retrieved large value of $f_\mathrm{M}=0.72_{-0.14}^{+0.15}$, which suggests a relatively homogeneous vertical aerosol distribution rather than a localised cloud deck. Compared to WASP-94~A~b, where particle sizes of $0.1$--$1~\mu$m were inferred \citep{Mukherjee2026}, we derive comparable or smaller particle sizes, potentially indicating less efficient particle growth or lofting.

Independent observations further support this picture. The optical \textit{Kepler} phase curve peaks after secondary eclipse \citep{Angerhausen2015, Esteves2015}, indicating enhanced reflectivity on the western (morning) hemisphere. This is naturally explained by a higher cloud coverage on the morning side, consistent with our transmission spectroscopy results.

The observed aerosol asymmetry also provides insight into their origin. Because aerosols are detected only on the morning limb, they are unlikely to consist primarily of photochemical hazes produced on the dayside. Although photochemical models predict a range of haze abundances depending on the atmospheric structure, they do not produce hazier morning limbs than evening limbs \citep{Mak2025}. Instead, our results are naturally explained if condensates form on the nightside, are advected to the morning terminator, and subsequently evaporate as they circulate across the dayside \citep{Mukherjee2026}. Extending limb-resolved observations across a broader range of planetary temperatures and gravities, while simultaneously constraining particle sizes, will provide powerful new tests of cloud microphysics and atmospheric circulation in hot Jupiter atmospheres.

\subsection{Is CO$_2$ enhanced on the morning limb?}

Although we do not find strong evidence for chemical asymmetries overall, the retrievals systematically allow for higher abundances on the morning limb. This is especially pronounced for CO$_2$, where on the evening limb we obtain only an upper limit of $\log (X_{\mathrm{CO}2,\,\mathrm{E}}) < -8.1$, whereas the morning-limb abundance is constrained to $\log (X_{\mathrm{CO}2,\,\mathrm{M}}) = -4.89_{-0.97}^{+0.91}$. A similar asymmetry has been reported for WASP-107~b \citep{Murphy2025}, with $\log (X_{\mathrm{CO}2,\,\mathrm{M}}) \approx -3.6$ and $\log (X_{\mathrm{CO}_2,\,\mathrm{E}}) \approx -5.2$. Such limb-to-limb differences in CO$_2$ abundance are qualitatively consistent with expectations from photochemistry \citep[][although we note that these works do not consider sub-solar metallicities.]{Tsai2023, Zamyatina2024}. However, models generally predict smaller contrasts than those inferred for Kepler-12~b and WASP-107~b. Given the limited precision, additional observations of Kepler-12~b are required to robustly establish the magnitude of this potential asymmetry. More broadly, limb-resolved spectroscopy across a wider sample of giant exoplanets will be essential to determine whether large CO$_2$ contrasts are common or not.

\subsection{Evidence for disequilibrium processes}

We find no evidence for CH$_4$ in the atmosphere of Kepler-12~b, despite the relatively low temperatures on the morning limb. Under chemical equilibrium, such temperatures would predict CH$_4$ abundances of around -6.5 to -4.5 between 10$^{-4}$ and 10$^{-1}$ bar, based on our elemental abundance constraints -- in clear tension with our morning-limb upper limit of $\log (X_{\mathrm{CH}_4,\,\mathrm{M}}) < -7.5$. This discrepancy provides evidence for disequilibrium processes, such as photochemistry or vertical and horizontal mixing, that could act to suppress the CH$_4$ abundance. Similar depletions of CH$_4$ relative to equilibrium expectations have been reported for WASP-39~b \citep{Constantinou2024, Lueber2024, Chen2025}, WASP-107~b \citep{Dyrek2024, Welbanks2024}, and HD 189733~b \citep{Fu2024}, based on recent JWST observations. In contrast, our upper limit on NH$_3$ is consistent with equilibrium predictions.

We also find that Na and K are strongly depleted relative to chemical equilibrium expectations, with 95\% upper limits of $<-8.5$ and $<-10.0$ for Na and K, respectively. These upper limits lie well below the log mixing ratios predicted for both the morning and evening limbs, of around -6.5 and -7.5 for Na and K, respectively, assuming that they scale with oxygen. This is consistent with similar findings for WASP-39~b, where Na and K are depleted by 2--3 dex relative the inferred metallicity based of C-, O-, and S-bearing species \citep{Constantinou2024}. A possible explanation is condensation and potential night-side cold trapping, which may remove alkali species from the observable atmosphere \citep{Parmentier2013, Kataria2016}.

\subsection{Possible evidence for atmospheric variability}

We find evidence for temporal variability in the transmission spectrum of Kepler-12~b based on two JWST observations obtained 71 days apart. As already indicated by the white-light analysis (see Section~\ref{sec:white_time}), the transit depth during the second visit is larger by $50_{-15}^{+16}$~ppm. By jointly fitting the spectra from both epochs, we are able to investigate the origin and significance of this difference.

The data strongly favour variability over a steady-state spectrum. We find that the observed changes can be reproduced by cloud variability, either through changes in cloud fraction or aerosol abundance. An alternative explanation is variability arising from stellar heterogeneities. Although this scenario is disfavoured by the Bayesian evidence, it nevertheless provides an acceptable fit to the limb-averaged observations and cannot be conclusively ruled out. Indeed, the observed signal may result from a combination of atmospheric and stellar effects. Overall, while the evidence for spectral variability is strong, its attribution to atmospheric or stellar variability remains more uncertain.

Evidence for atmospheric variability in transiting exoplanets remains scarce. Recent analyses of repeated HST observations of the ultra-hot Jupiter WASP-121~b have reported significantly larger variations in the dayside temperature structure between epochs \citep{Changeat2024}. Our results demonstrate the potential of repeated JWST transmission spectroscopy to probe temporal changes in exoplanet atmospheres and provide tentative evidence that the atmospheric structure of a transiting giant exoplanet may evolve on timescales of months.

\subsection{The importance of covariance in limb-resolved spectroscopy} \label{sec:atmo_covariance}

A key advance of this work is the explicit treatment of covariance. This requires both estimating the covariance matrix of the spectra and propagating it through the retrieval analysis. Two primary sources of covariance are present. First, there is a strong negative covariance between the inferred morning and evening transit depths at a given wavelength \citep{Espinoza2021}. Second, uncertainties in shared parameters, such as the orbital geometry and limb-darkening coefficients, introduce correlated structure across wavelengths and between the two limbs. Consequently, the relative shape of the spectra is generally much better constrained than the absolute offset between the morning and evening spectra. We show the covariance matrix of the morning and evening spectra in Appendix~\ref{app:cov}.

Accounting for covariance is essential for robust atmospheric inference of limb-resolved spectra, as shown in this work. If neglected, the retrievals treat the two limbs as independent measurements and fail to utilise the additional information present in the covariance. In our analysis (see Section~\ref{sec:asymmetric}), this leads to substantially weaker constraints: the uncertainty of the log volume-mixing ratio of H$_2$O increases by approximately a factor of two, while CO$_2$ and H$_2$S are no longer detected. The impact is also reflected in the Bayesian evidence, which decreases dramatically ($\ln B = -176$).

Our framework also addresses a common challenge in limb-resolved spectroscopy: the uncertain absolute offset between the morning and evening spectra \citep{Espinoza2024, Murphy2024a, Fu2025}. By propagating uncertainties in the shared model parameters, we construct a covariance term ($\mathbf{\Sigma}_\mathrm{common}$) that quantifies the uncertainty in the absolute transit depths. As a result, the absolute offset between the two spectra is no longer arbitrary, but instead becomes a well-defined quantity that is rigorously incorporated into the atmospheric retrievals.

\subsection{Future prospects}

The results presented here motivate several directions for future work. The pronounced cloud asymmetry and inferred particle sizes call for comparisons with cloud microphysics models. In particular, coupling cloud formation processes -- such as nucleation, growth, and settling -- with atmospheric dynamics will be crucial for understanding the persistence of a cloudy morning limb alongside a largely clear evening limb at observable pressures. The large temperature contrast between the two limbs also provides an important constraints for general circulation models. Current models generally predict smaller terminator temperature contrasts, suggesting that key physical processes, such as atmospheric drag, radiative timescales, or cloud--radiation feedback, may need to be explored further to reproduce the observed behaviour. The possible evidence for temporal variability provides additional motivation to investigate the atmospheric dynamics underlying these differences. Detailed comparisons with disequilibrium chemistry models, including vertical and horizontal mixing, photochemistry, and condensation, will be essential for interpreting the apparent depletion of CH$_4$ and alkali species in Kepler-12~b, as well as the tentative evidence for enhanced CO$_2$ on the morning limb. Future observations will provide valuable additional constraints. In particular, upcoming JWST/MIRI LRS observations (GO Program 7849, PI: S. Moran \& L. Alderson) will extend the spectral coverage into the mid-infrared, potentially enabling improved constraints on the aerosol composition.

Taken together, these results highlight the power of multi-epoch, limb-resolved transmission spectroscopy. With JWST, we are no longer limited to static, one-dimensional views of exoplanet atmospheres, but can begin to map their spatial structure and temporal evolution via transmission spectroscopy. Kepler-12~b provides a compelling example of this emerging frontier, revealing how atmospheric dynamics, chemistry, and cloud formation interact to shape the observable properties of giant exoplanets.

\begin{acknowledgments}
This work is based on observations made with the NASA/ESA/CSA JWST. The data were obtained from the Mikulski Archive for Space Telescopes at the Space Telescope Science Institute, which is operated by the Association of Universities for Research in Astronomy, Inc., under NASA contract NAS 5-03127 for JWST. These observations are associated with JWST GO Program 3969 (PIs: N.~Espinoza \& D.~Powell). Support for program JWST GO Program 3969 was provided by NASA through a grant from the Space Telescope Science Institute. MH and DS acknowledges funding as part of JWST GO program 3969. MS and MCN acknowledges support from the Heising-Simons Foundation through the 51 Pegasi b fellowship. DAC is supported by the Max Planck Society. R.B. acknowledges support from FONDECYT Project 1241963. NJM acknowledges support via a UKRI Future Leaders Fellowship and extension [grant numbers MR/T040866/1 \& MR/Z000122/1], a Science and Technology Facilities Council Consolidated Grant [ST/R000395/1] and the Leverhulme Trust through a research project grant [RPG-2020-82].
\end{acknowledgments}

\section*{Data Availability}  The JWST data presented in this article were obtained from the Mikulski Archive for Space Telescopes (MAST) at the Space Telescope Science Institute. The specific observations analyzed can be accessed via doi:\href{https://doi.org/10.17909/zetv-mj35}{10.17909/zetv-mj35}. The transmission spectra of Kepler-12~b reported in this work are available on the Open Science Framework at doi:\href{https://doi.org/10.17605/OSF.IO/T654A}{10.17605/OSF.IO/T654A}.





%
\facilities{JWST (NIRSpec)}



\appendix

\section{Treatment of a modelling issue in \texttt{catwoman}} \label{app:catwoman}

We identified an issue in the implementation of \texttt{catwoman} \citep[before version 1.1.0;][]{Espinoza2021, Jones2022} that significantly impacted the inferred limb-resolved transmission spectra of Kepler-12~b. The issue originates from the computation of the angle, denoted as $\theta$ in \cite{Espinoza2021}, between the line that separates the two semicircles and the line extending from the centre of the planet to the centre of the star. This computation depends on the side of the star on which the planet is located, determined by the condition $t > T_0$ (as specified in Table A.2 of \cite{Espinoza2021}). However, as initially implemented, this approach presents two main problems: first, the value of $T_0$ used for this condition is fixed at model initiation and does not get updated if $T_0$ is later changed. Second, the method fails when the value of $T_0$ corresponds to an epoch different from the data, as it does not account for periodicity when evaluating this condition. Consequently, depending on how \texttt{catwoman} is used, the computation of the geometry may become inaccurate. To resolve these issues, we replaced the $t > T_0$ condition with $x_\mathrm{p} > 0$, where $x_\mathrm{p}$ is the x-position of the planet relative to the centre of the star. This change resolved the issues and is now implemented in the latest version of \texttt{catwoman} (version 1.1.0). 

\section{Noise characterisation} \label{app:noise}

\begin{figure}
\centering
	\includegraphics[width=0.49\textwidth]{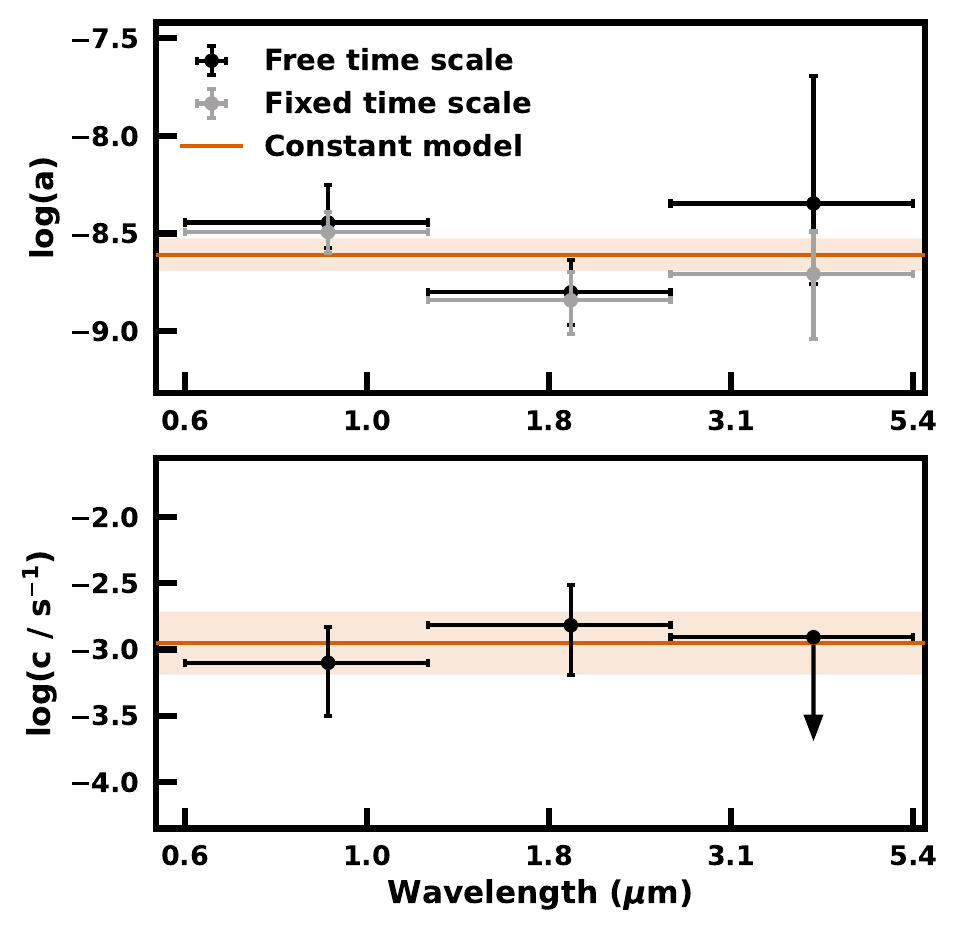}
    \vspace{-5mm}
    \caption{Wavelength dependence of the GP kernel parameters. The top and bottom panels show the inferred GP amplitudes and timescales, respectively. The grey and black data points were obtained assuming fixed and free timescales, respectively. The orange lines represent the median models, as described in Appendix~\ref{app:noise}, along with the 1$\sigma$ intervals in a lighter shade of orange.
    }
    \label{fig:gp_params} 
\end{figure}

As mentioned in Section~\ref{sec:orbital_properties}, we find strong evidence for time-correlated noise in the light curves, which could be due to instrumental systematics, stellar granulations, or both. Here, we characterise the wavelength dependence of this noise -- finding it approximately wavelength independent at the current precision. To do this, we split the data into three large bins with $R \approx 1$ and fit these light curves independently, while fixing the orbital parameters to those of Table~\ref{tab:wlc_params}, assuming no limb asymmetry. Using an exponential kernel, defined as $\kappa(\Delta t) = a e^{-c \Delta t}$, we find that the GP amplitudes (i.e., variance) and time-scales are consistent across the wavelength bins, as illustrated in Figure~\ref{fig:gp_params}. Note that the timescale is not well-constrained for the reddest data point (3.1-5.4~$\mu$m); however, we find that the 95\% upper limit is consistent with the data at shorter wavelengths. Therefore, to estimate the average of the timescale, we only used the two bins in the 0.6-3.1~$\mu$m range. Similarly, since the amplitude and timescale are strongly correlated, we fixed the GP timescale to the above average value and re-fit the light curves to obtain our final estimate of the amplitudes. As a result, we find average values of $\log(a) = -8.61\pm0.08$ and $\log(c / \mathrm{s}^{-1}) = -2.96\pm0.24$, which we adopt for the spectroscopic light curve fitting.

\section{Stellar limb darkening} \label{app:limb-darkening}

\begin{figure}
\centering
	\includegraphics[width=0.49\textwidth]{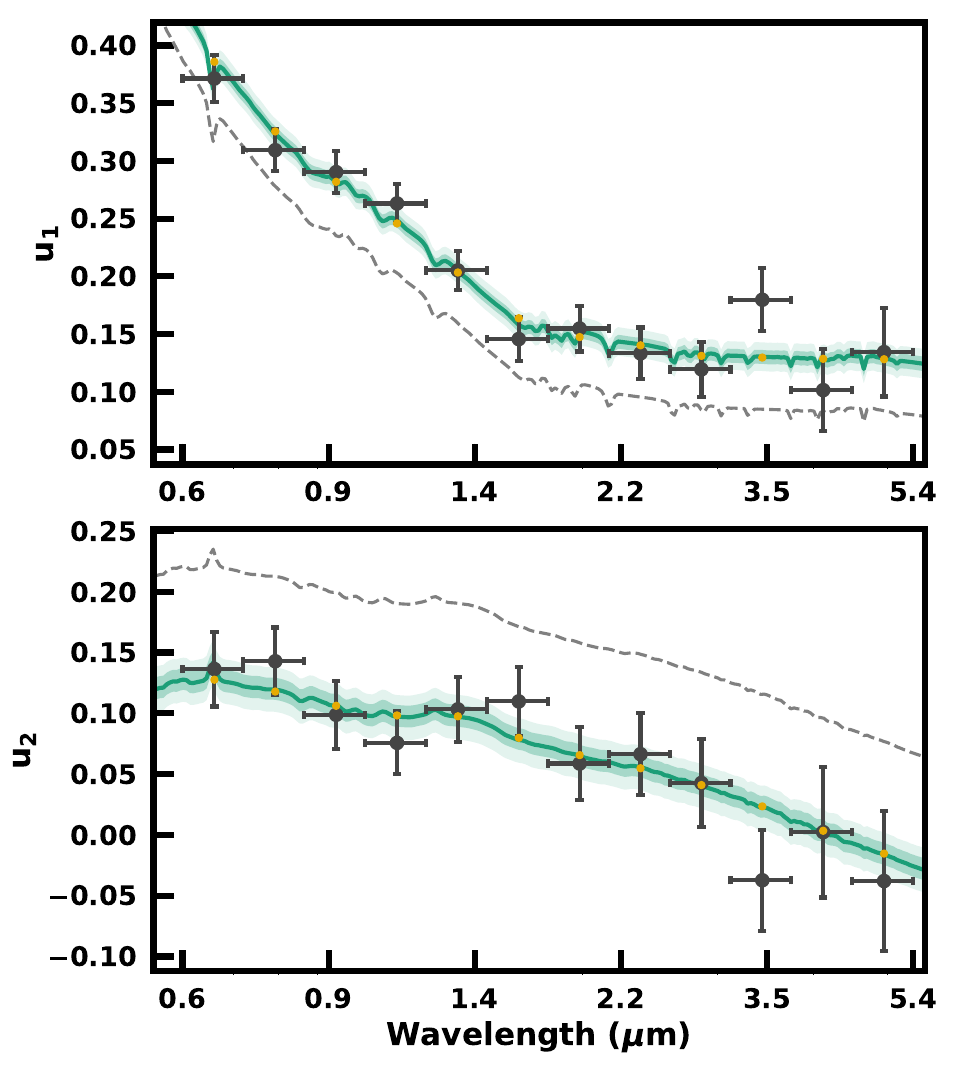}
    \vspace{-5mm}
    \caption{Stellar limb-darkening coefficients (LDCs) of Kepler-12 in the NIRSpec PRISM bandpass. The black data points correspond to the empirical LDCs at $R = 5$. The green curves represent the median model along with the 1$\sigma$ and 2$\sigma$ contours in lighter shades of green. The yellow points correspond to the binned model. The dashed curves show the median model without any offsets. The model is described in Appendix~\ref{app:limb-darkening}.}
    \label{fig:ld} 
\end{figure}

Stellar limb darkening is an important consideration for transmission spectroscopy \citep[e.g.,][]{Espinoza2015}. Several approaches have been used in the literature when it comes to JWST transmission spectroscopy, ranging from interdependently fitting the limb-darkening coefficients (LDCs) for each light curve to using values predicted by stellar models \citep[either directly or as priors, see e.g.,][]{Rustamkulov2023, Alderson2023, Feinstein2023}. In this work, we adopt a hybrid approach where we fit a global model to the empirical LDCs.

First, we bin the light curves to a resolution of $R = 5$ and fit these using the the system properties in Table~\ref{tab:wlc_params}, assuming no limb-asymmetry. As for the analysis of the white light curves, we utilise quadratic LDCs using the parametrisation by \cite{Kipping2013} and fitted for an exponential trend using the parametrisation described in Appendix~\ref{app:trend}. We jointly fitted both visits, used the wavelength-independent GP parameters derived in Appendix~\ref{app:noise}, and fitted for a white noise inflation parameter. Next, we construct a model of the LDCs using the MPS-ATLAS stellar model grid \citep[MPS1; ][]{Kostogryz2022, Kostogryz2024}. For each wavelength bin, we generate a synthetic (noiseless) light curve with \texttt{spotrod}\footnote{We employ \texttt{spotrod} to numerically compute transit light curves using custom stellar intensity profiles.} \citep{spotrod2014} using the intensity profile from the stellar model (while taking into account the spectral resolution and throughput function of NIRSpec PRISM), as a function of metallicity~[Fe/H], surface gravity~$\log (g)$ and effective temperature~T$_\mathrm{eff}$. We then fit these light curves to obtain the model LDCs \citep{Howarth2011}, which we repeat across a grid of $\mathrm{[Fe/H]} \in [-0.3, 0.3]$, $\log (g) \in [4.0, 5.0]$ and T$_\mathrm{eff} \in [5500~\mathrm{K}, 6500~\mathrm{K}]$. 

We then use the empirical LDCs in combination with the above model to fit for the stellar properties (three parameters), wavelength-independent offsets for $u_1$ and $u_2$ (two parameters), and an error inflation parameter (one parameter). We allow for potential offsets since previous work have suggested that these are needed to fit similar JWST data \citep[e.g.,][]{Rustamkulov2023}. For each model evaluation, we linearly interpolate the model LDCs to the desired stellar properties. We construct a likelihood function that accounts for the covariance between $u_1$ and $u_2$ and use \texttt{MultiNest} to sample from the posterior distribution. We adopt [Fe/H]$\,\sim N(0.01, 0.05)$, $\log (g) \sim N(4.17, 0.02)$ as informative priors, based on stellar parameters derived from spectroscopy (as described in Appendix~\ref{app:stellar}). For the temperature, we first adopt T$_\mathrm{eff} \sim U(5500~\mathrm{K}, 6500~\mathrm{K})$. This results in a posterior constraint of T$_\mathrm{eff} = 6101 \pm 91$~K, which is in good agreement with the value of $6100\pm100~\mathrm{K}$ derived from spectroscopy. For the final fit, we adopt T$_\mathrm{eff} \sim N(6100~\mathrm{K}, 100~\mathrm{K})$ as the prior for the temperature to be consistent with the treatment of the other parameters. 

As shown in Figure~\ref{fig:ld}, the resulting model fits the data well. We find that significant offsets are needed, both in $u_1$ and $u_2$, suggesting that the wavelength-dependence of the LDCs is well described by the stellar model, while the average is not. The inferred offsets of $0.0457 \pm 0.0079$ and $-0.0931 \pm 0.0098$ for $u_1$ and $u_2$, respectively, indicate that the stellar limb of Kepler-12 is brighter than predicted by the stellar models, similar to WASP-39 \citep{Rustamkulov2023}. For robustness, we also adopt the ATLAS9 stellar model grid \citep{Kurucz1993} and find that it provides a comparable model fit ($|\ln B| = 0.2$) and a consistent temperature estimate ($6190 \pm 86$~K, using broad priors). Finally, with these fitted parameters in hand, we generate model LDCs at native resolution for the spectroscopic light curve fitting. In summary, this framework not only enables us to determine precise LDCs, but it also demonstrates that it is possible to independently infer stellar effective temperatures from the wavelength dependence of the LDCs using high-quality JWST data. Future work is needed to assess whether this is applicable to other stellar types.

\section{Treatment of light-curve trends} \label{app:trend}

We utilise the following parametrisation to model the exponential trends in the light curves:
\begin{equation} \label{eq:trend}
\begin{aligned}
    F_\mathrm{obs}(t) &= F_\star \left( (1 + A) (1- w) + w\right)\,, \\[0.3em] w &= \frac{1 - e^{-(t-t_{\mathrm{start}})/\tau}}{1 - e^{-(t_\mathrm{end}-t_{\mathrm{start}})/\tau}}\,,
\end{aligned}
\end{equation}
where $F_\mathrm{obs}(t_\mathrm{start}) = F_\star \, (1 + A)$ and $F_\mathrm{obs}(t_\mathrm{end}) = F_\star$ are the fluxes at the beginning and end of the observation, respectively, independent of the timescale $\tau$. Moreover, for large values of $\tau$, we have $w(\tau \to \infty) = (t - t_\mathrm{start}) / (t_\mathrm{end} - t_\mathrm{start})$, such that equation~\eqref{eq:trend} reduces to a straight line between the two endpoints. This parametrisation provides a natural way to interpolate between an exponential and a linear function while avoiding strong degeneracies between the parameters.

\begin{figure}
\centering
	\includegraphics[width=0.49\textwidth]{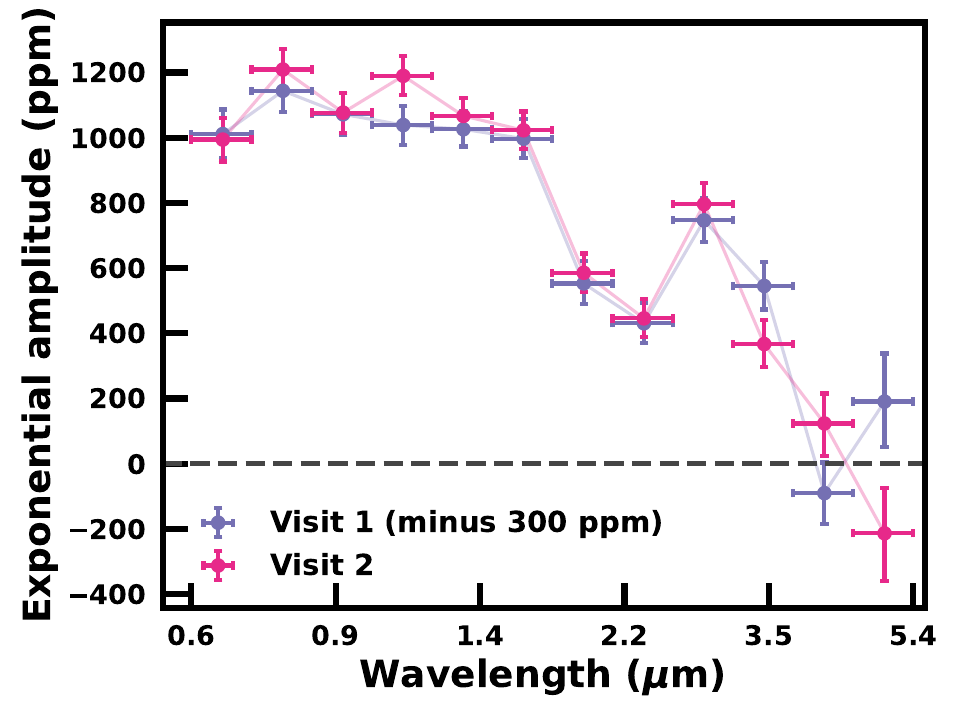}
    \caption{ Amplitudes of the light-curve exponential settling trends as a function of wavelength and visit. After applying an offset of -300~ppm, the wavelength dependence is consistent between the two visits. }
    \label{fig:exp_amp} 
\end{figure}

Next, we characterise the wavelength dependence of the exponential trends in the $R=5$ light curves (see Appendix~\ref{app:limb-darkening}) to inform the fitting of the spectroscopic light curves. As shown in Figure~\ref{fig:exp_amp}, we find that the amplitudes of the exponential trends are strongest at short wavelengths, reaching a level of around 0.1\%. Furthermore, apart from an overall offset in the amplitudes, the wavelength dependence appears to be consistent between the two visits. The fact that we observe these exponential trends in two independent visits, with similar characteristics, may suggests that the cause is instrumental. Previous works have also invoked exponential trends to explain the light curves of NIRSpec time-series observations \citep{Rustamkulov2023, Cassese2026, Kipping2026}, further indicating a systematic effect. 

\begin{figure}
\centering
	\includegraphics[width=0.49\textwidth]{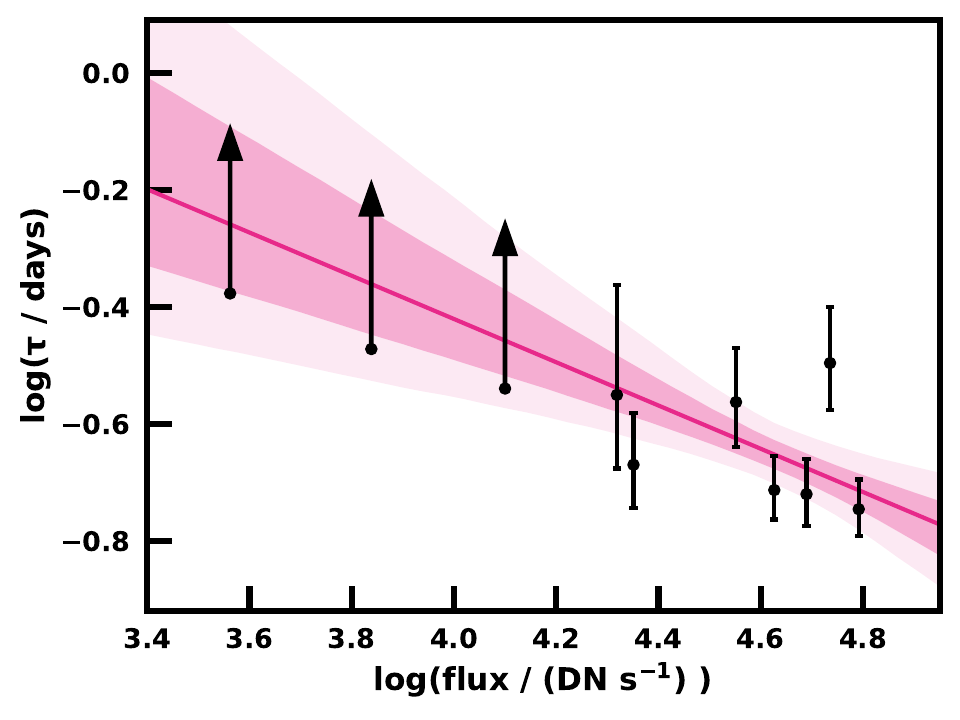}
    \caption{Exponential settling timescale of the light curves as a function of flux. For reference, flux lower than $10^4$~DN/s corresponds to wavelengths longer than 2.5~$\mu$m.  }
    \label{fig:exp_timescale} 
\end{figure}

Detector settling is a possible explanation for the observed trends, similar to what has been reported for MIRI \citep[e.g.,][]{Bouwman2023, Bell2024}. Recent investigations into the detector settling behaviour of MIRI have found evidence that the settling timescale depends on the flux incident on the pixels \citep{Connors2025, Holmberg2026}. For this reason, we investigate this potential dependence in our NIRSpec observations. Figure \ref{fig:exp_timescale} shows the measured timescales as a function of the mean flux of each wavelength bin. We find that regions of high flux, corresponding to short wavelengths, have shorter exponential timescales, while at lower fluxes we derive only limits. Using these empirical constraints, we fit a linear relation using the form
\begin{equation}
    \log(\tau) = p_1  \log(F_\lambda) + p_2\,,
\end{equation}
where $\tau$ is the timescale (in days), $F_\lambda$ is the flux (in DN/s), and $p_1$ and $p_2$ are dimensionless constants. We fit this relationship to the data by constructing an empirical likelihood function using the posterior samples of the exponential timescales for each bin. We take this approach because the marginal posterior distributions are not normally distributed. The resulting model fit is displayed in Figure~\ref{fig:exp_timescale}. In contrast, when we fit only a constant value (i.e., $p_1 = 0$), the fit is poorer ($\ln B = -2.7$). With the inferred constraints on $p_1$ and $p_2$, we proceed to predict the exponential timescale at native resolution, which we then utilise to fit the spectroscopic light curves. For added robustness, we also analyse the NIRSpec PRISM observation of HAT-P-65~b within our program (JWST GO 3969) and find a similar correlation between the exponential timescale and flux. 

\begin{figure}
\centering
	\includegraphics[width=0.49\textwidth]{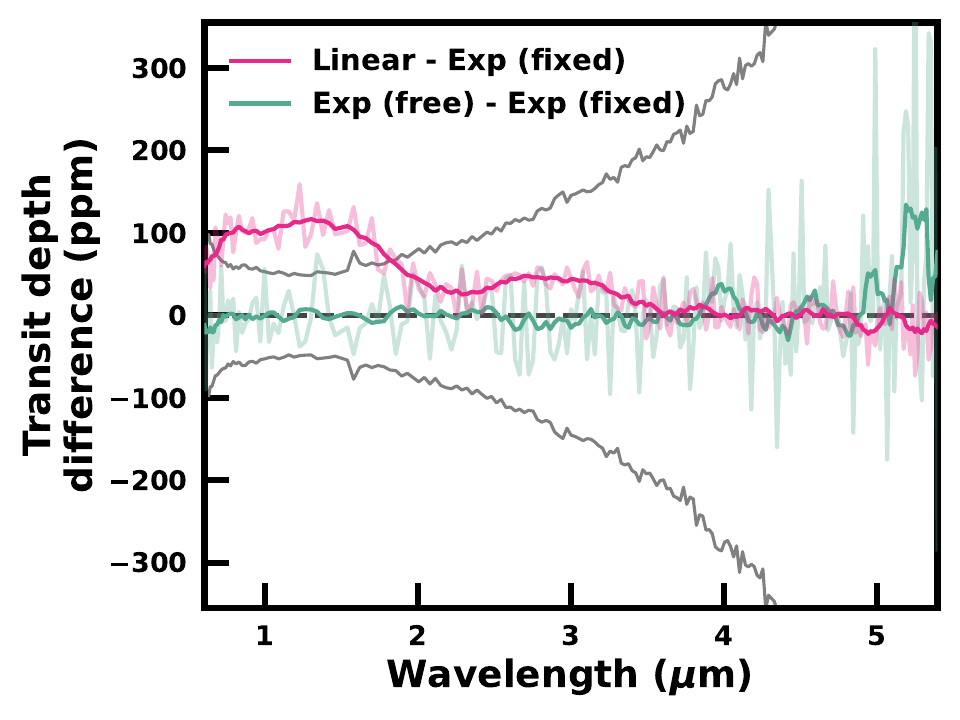}
        \vspace{-5mm}
    \caption{ Difference between transmission spectra from the second visit derived under different trend assumptions. The green curves show the difference between spectra obtained when the exponential timescale is treated as a free parameter and when it is fixed to the relation derived in Appendix~\ref{app:trend}. The pink curves show the difference between spectra obtained using an exponential trend and a linear trend. The faint curves indicate the raw differences, while the opaque curves show a smoothed version for clarity. The grey curves indicate the 1$\sigma$ uncertainties. }
    \label{fig:trend_assumptions} 
\end{figure}

Finally, we assess the impact of imposing the above timescale–flux relation on the inferred transmission spectrum. We compare the nominal limb-averaged spectrum, which adopts this relation, to a version in which the timescale is treated as a free parameter. For this comparison, we use the second visit. The difference between the spectra is shown in Figure~\ref{fig:trend_assumptions} (green curve). We find no systematic differences, with most data points consistent within uncertainties. We therefore conclude that fixing the exponential timescale according to the above relation does not bias the inferred transmission spectrum. The advantage of this approach is improved precision: the spectrum is $\sim$15\% more precise below 1.7~$\mu$m, with the gain decreasing toward longer wavelengths. In contrast, adopting a linear trend significantly biases the transit depth at wavelengths shorter than $\sim$3.5~$\mu$m, by up to $\sim$100 ppm at 1~$\mu$m (see the pink curve in Figure~\ref{fig:trend_assumptions}). Such biases can affect inferences of atmospheric scattering slopes or stellar heterogeneity.

\section{Stellar Properties of Kepler-12} \label{app:stellar}

We computed the atmospheric stellar parameters for Kepler-12 using publicly available HIRES spectra. We downloaded the raw HIRES data associated with Kepler-12 obtained on 2009-09-05. We processed the data with the \texttt{ceres} pipeline \citep{ceres} to generate the wavelength calibrated and continumm normalised spectrum. Then we used the \texttt{zaspe} code \citep{zaspe} to obtain the atmospheric parameters from the high resolution spectrum. \texttt{zaspe} uses synthetic spectra generated with the ATLAS9 models \citep{kurucz1979, castelli2003} that are compared to the observed spectrum in the spectral regions that are the most sensitive ones to changes in the stellar parameters. Reliable uncertainties in the atmospheric parameters are computed by taking into account that the main error budget originates from systematic mismatches between the observed spectrum and the best fit model. We provide the derived properties in Table~\ref{tab:stellar_params}.

\begin{table}
\renewcommand{\arraystretch}{1.55} 
\setlength{\tabcolsep}{14pt}
\centering
\begin{tabular}{lc}
\hline \hline
Parameter & Value \\ \hline
T$_\mathrm{eff}$    (K) &  $6100 \pm 100$ \\ 
$[$Fe/H$]$  (dex) & $0.01 \pm 0.05$ \\
$v \sin(i)$ (km/s) & $3.3 \pm 0.5$ \\
$\log(g)$   (dex) & $4.17 \pm 0.02$ \\
M$_\star$   (M$_\odot$) & $1.13 \pm 0.04$ \\
R$_\star$   (R$_\odot$) & $1.45 \pm 0.02$ \\ \hline
\end{tabular}
\vspace{1mm}
\caption{Stellar properties of Kepler-12 derived in this work using HIRES data.
}
\label{tab:stellar_params}
\end{table}

\section{Covariance Matrix} \label{app:cov}

\begin{figure}
\centering
	\includegraphics[width=0.49\textwidth]{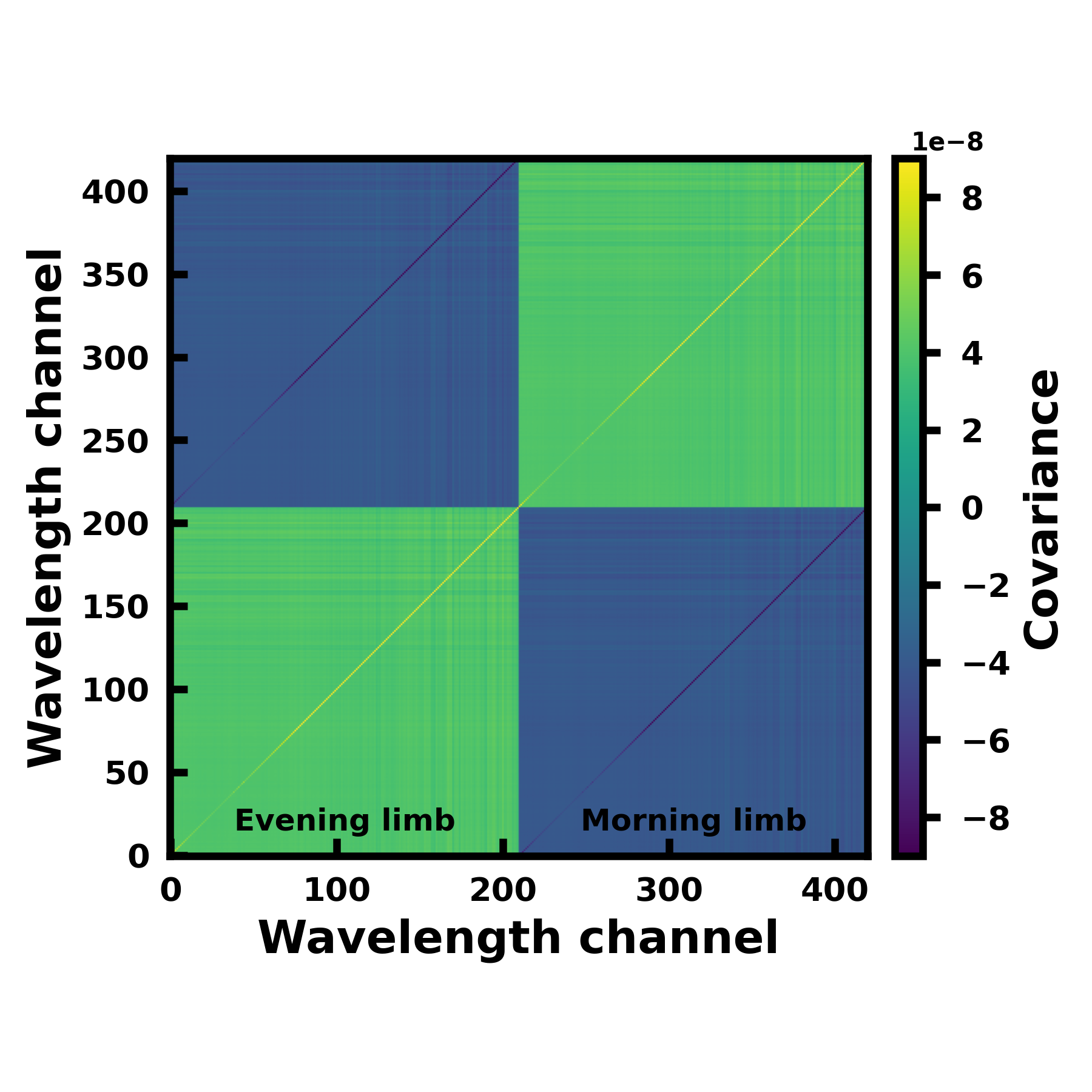}
     \vspace{-12mm}
    \caption{ The full covariance matrix of the morning and evening spectra at native spectral resolution (binned over every two detector pixels), corresponding to the average spectra from both visits. The first 210 spectral channels correspond to the evening spectrum, and the remaining 210 channels correspond to the morning spectrum. Within each quadrant, the off-diagonal covariance is approximately constant at \(\pm\, 4\times10^{-8}\), corresponding to a shared baseline uncertainty of approximately 200 ppm for each spectrum. This is the covariance matrix used for the retrievals in Section~\ref{sec:asymmetric}.
    }
    \label{fig:cov} 
\end{figure}

Figure~\ref{fig:cov} shows the covariance matrix for the time-averaged morning and evening spectra. The construction of this matrix is described in Section~\ref{sec:spectrum}, which is then utilised for the limb-resolved retrievals outlined in Section~\ref{sec:asymmetric}.

\section{A Computationally Optimised Forward Model} \label{app:model_implementation}

We optimised the implementation of the \texttt{POSEIDON} \citep{MacDonald2017, MacDonald2022, MacDonald2024} transmission-spectrum forward model to improve computational performance. To leverage GPU acceleration, we implemented the majority of the forward-model calculations using \texttt{MLX} \citep{mlx2023}. After generating the atmospheric profiles and setting up the geometry, the calculations of extinction coefficients, radiative transfer, spectral convolution, and binning are performed in a fully vectorised manner. We also optimised the interpolation of aerosol cross sections. Retrievals performed with this optimised version reproduce the results obtained with the standard \texttt{POSEIDON} implementation. These optimisations make repeated forward-model evaluations more computationally tractable for the limb-resolved retrieval framework developed in this work. For a representative model containing 100 atmospheric layers, 11 gas species, patchy Mie-scattering aerosols, and 47,825 wavelength points ($R=20{,}000$ between 0.5--5.5~$\mu$m), the optimised implementation requires 5.5~ms per forward-model evaluation on an Apple M3 Max chip with 40 GPU cores and 64 GB of unified memory. Under the same conditions, the standard \texttt{POSEIDON} implementation requires 246.1~ms when executed on a single CPU thread.

In addition to the optimisations mentioned above, we also addressed an issue related to the implementation of patchy clouds in version 1.3.2 of \texttt{POSEIDON}. In the standard implementation, the cloud fraction $\phi$ and the parameter $f$, which determines the vertical aerosol profile, were mapped to the same value. As a result, $f$ was set internally to the value of $\phi$, while the input value of $f$ was left unused. By resolving this, we are able to retrieve both parameters.

\section{System parameters} \label{app:system_parameters}

Table \ref{tab:wlc_params} presents the Kepler-12~b system parameters derived from the white light curve analysis of both visits (see Section~\ref{sec:orbital_properties}), along with the adopted priors. We consider two standard white light curve models: one including limb asymmetries and one without. All parameters are consistent between the two models.

\begin{table*}
\renewcommand{\arraystretch}{1.55}
\setlength{\tabcolsep}{14pt}
\centering
\begin{tabular}{lcccc}
\hline \hline
Parameter & Prior & Value (limb-averaged) & Value (limb-resolved)  \\ \hline
P (days) & N(4.4379637, 0.0000002) & $4.437963747_{-0.000000022}^{+0.000000022}$ & $4.43796370_{-0.00000016}^{+0.00000016}$ \\
T$_0$ (BJD) & N(55003.50835, 0.00002) &  $55003.508350_{-0.000019}^{+0.000019}$ & -  \\ 
 & U(55003.50735, 55003.50935) &  - & $55003.50859_{-0.00028}^{+0.00027}$  \\ 
$\sqrt{e} \cos \omega$ & N(-0.001, 0.054) & $0.000_{-0.051}^{+0.049}$ & $-0.001_{-0.048}^{+0.050}$ \\
$\sqrt{e} \sin \omega$ & N(0.001, 0.114) & $-0.006_{-0.067}^{+0.068}$ & $-0.004_{-0.065}^{+0.067}$ \\
$b$ & N(0.174, 0.011) & $0.1797_{-0.0081}^{+0.0081}$ & $0.1803_{-0.0082}^{+0.0079}$ \\
$\rho_\star$ ($\rho_\odot$) & N(0.354, 0.017) & $0.3503_{-0.0056}^{+0.0064}$ & $0.3501_{-0.0055}^{+0.0058}$ \\
$q_1$ & U(0, 1) & $0.1059_{-0.0058}^{+0.0064}$ & $0.1061_{-0.0058}^{+0.0062}$  \\ 
$q_2$ & U(0, 1)  & $0.353_{-0.023}^{+0.023}$ & $0.352_{-0.022}^{+0.022}$  \\
$(R_\mathrm{p} / R_\star)_1$ & U(0.108, 0.126) & $0.118933_{-0.000084}^{+0.000085}$ & $0.118929_{-0.000080}^{+0.000079}$$^\dagger$ \\ 
$(R_\mathrm{p} / R_\star)_2$ & U(0.108, 0.126) & $0.119135_{-0.000081}^{+0.000081}$ & $0.119140_{-0.000077}^{+0.000079}$$^\dagger$ \\ 
$(R_\mathrm{p,\,E} / R_\star)_1$ & U(0.108, 0.126) & - & $0.1171_{-0.0017}^{+0.0018}$ \\ 
$(R_\mathrm{p, \,E} / R_\star)_2$ & U(0.108, 0.126) & - & $0.1180_{-0.0017}^{+0.0017}$ \\ 
$(R_\mathrm{p, \,M} / R_\star)_1$ & U(0.108, 0.126) & - & $0.1208_{-0.0017}^{+0.0016}$ \\ 
$(R_\mathrm{p, \,M} / R_\star)_2$ & U(0.108, 0.126) & - & $0.1203_{-0.0017}^{+0.0017}$ \\ 
$i$ (deg) & - &  $88.715_{-0.060}^{+0.062}$$^\dagger$ & $88.709_{-0.060}^{+0.063}$$^\dagger$ \\ 
$a / R_\star$ & - & $8.008_{-0.043}^{+0.048}$$^\dagger$ & $8.007_{-0.042}^{+0.044}$$^\dagger$   \\ 
$u_1$ & - & $0.2298_{-0.0091}^{+0.0087}$$^\dagger$ & $0.2296_{-0.0087}^{+0.0086}$$^\dagger$   \\ 
$u_2$ & - & $0.096_{-0.017}^{+0.018}$$^\dagger$ & $0.096_{-0.017}^{+0.017}$$^\dagger$  \\ \hline
\end{tabular}
\vspace{1mm}
\caption{Inferred system parameters of Kepler-12~b using our two NIRSpec PRISM visits. The prior distributions on the period $P$, mid-transit time T$_0$, impact parameter $b$, stellar density $\rho_\star$, $\sqrt{e} \cos \omega$, and $\sqrt{e} \sin \omega$ are obtained from \cite{Fortney2011}. The prior distribution on the planet-to-star radius ratios $(R_\mathrm{p} / R_\star)_i$ corresponds to the Kepler transit depth~$\pm 2000$~ppm. The limb-darkening parameters $q_1$ and $q_2$ represent the quadratic limb-darkening parametrisation by \cite{Kipping2013}. We give T$_0$ in terms of BJD $-$ 2400000.5 days. Derived parameters are denoted with $^\dagger$.
}
\label{tab:wlc_params}
\end{table*}

\section{Atmospheric retrieval constraints} \label{app:retrieval_constraints}

We present the full set of priors and parameter estimates from the atmospheric retrieval analysis of Kepler-12~b in Tables~\ref{tab:normal_atm_params} and \ref{tab:asymmetric_atm_params}. Details of the retrieval analysis are provided in Section~\ref{sec:retrievals}.

\begin{table*}
\renewcommand{\arraystretch}{1.55} 
\setlength{\tabcolsep}{14pt}
\centering
\begin{tabular}{lccccc}
\hline \hline
Parameter & Prior & Visit 1 & Visit 2 & Combined  \\ \hline
$\log (\mathrm{P}_\mathrm{ref}/\mathrm{bar})$ & $\mathcal{U}$(-6, 0) & $-2.30_{-0.17}^{+0.17}$ & $-1.77_{-0.19}^{+0.16}$ & $-2.09_{-0.13}^{+0.12}$ \\
T$_\mathrm{iso}$ (K) & U(450, 2000) &  $1237_{-105}^{+93}$ & $1290_{-110}^{+100}$ & $1346_{-74}^{+69}$ \\ 
$\log (\mathrm{X}_{\mathrm{H}_2\mathrm{O}})$ & $\mathcal{U}$(-12, -0.3) & $-4.04_{-0.16}^{+0.20}$ & $-4.52_{-0.13}^{+0.14}$ & $-4.39_{-0.11}^{+0.12}$ \\
$\log (\mathrm{X}_{\mathrm{CO}_2})$ & $\mathcal{U}$(-12, -0.3) & $-7.06_{-0.31}^{+0.27}$ & $-8.54_{-1.31}^{+0.52}$ & $-7.61_{-0.22}^{+0.20}$ \\
$\log (\mathrm{X}_{\mathrm{H}_2\mathrm{S}})$ & $\mathcal{U}$(-12, -0.3) & $-4.57_{-0.22}^{+0.21}$ & $< -5.7$ & $-5.09_{-0.21}^{+0.18}$ \\
$\log (\mathrm{X}_{\mathrm{CO}})$ & $\mathcal{U}$(-12, -0.3) & $< -4.7$ & $< -5.2$ & $< -4.8$ \\
$\log (\mathrm{X}_{\mathrm{CH}_4})$ & $\mathcal{U}$(-12, -0.3) & $< -8.0$ & $< -8.3$ & $< -8.3$ \\
$\log (\mathrm{X}_{\mathrm{NH}_3})$ & $\mathcal{U}$(-12, -0.3) & $< -6.6$ & $< -7.0$ & $< -6.8$ \\
$\log (\mathrm{X}_{\mathrm{HCN}})$ & $\mathcal{U}$(-12, -0.3) & $< -7.3$ & $< -8.1$ & $< -8.0$ \\
$\log (\mathrm{X}_{\mathrm{C}_2\mathrm{H}_2})$ & $\mathcal{U}$(-12, -0.3) & $< -7.7$ & $< -8.2$ & $< -8.1$ \\
$\log (\mathrm{X}_{\mathrm{SO}_2})$ & $\mathcal{U}$(-12, -0.3) & $< -6.6$ & $< -7.5$ & $< -7.1$ \\
$\log (\mathrm{X}_{\mathrm{Na}})$ & $\mathcal{U}$(-12, -0.3) & $< -6.9$ & $< -7.9$ & $< -8.4$ \\
$\log (\mathrm{X}_{\mathrm{K}})$ & $\mathcal{U}$(-12, -0.3) & $< -9.2$ & $< -10.2$ & $< -10.1$ \\
$\log (\mathrm{P}_\mathrm{deck}/\mathrm{bar})$ & $\mathcal{U}$(-3, 2) & $-0.5_{-1.3}^{+1.5}$ & $-0.2_{-1.1}^{+1.2}$ & $-0.1_{-1.3}^{+1.3}$ \\
$\log(r_{\mathrm{m}}/\mu \mathrm{m})$ & $\mathcal{U}$(-3, 1) & $-1.36_{-0.29}^{+0.34}$ & $-1.05_{-0.42}^{+0.31}$ & $-0.98_{-0.33}^{+0.21}$ \\
$\log(n_{\mathrm{max,\,MgSiO}_3})$ & $\mathcal{U}$(0, 25) & $12.6_{-3.1}^{+3.7}$ & $12.6_{-4.3}^{+6.0}$ & $10.6_{-2.6}^{+4.7}$ \\
$\phi$ & $\mathcal{U}$(0, 1) & $0.379_{-0.062}^{+0.078}$ & $0.480_{-0.067}^{+0.086}$ & $0.414_{-0.043}^{+0.045}$\\
$f$ & $\mathcal{U}$(0, 1) & $0.54_{-0.14}^{+0.18}$ & $0.32_{-0.10}^{+0.19}$ & $0.54_{-0.16}^{+0.23}$ \\ \hline
\end{tabular}
\vspace{1mm}
\caption{Retrieved atmospheric parameters for the limb-averaged analysis of Kepler-12~b, as described in Sections~\ref{sec:ret:normal} and \ref{sec:time-resolved}.
}
\label{tab:normal_atm_params}
\end{table*}

\begin{table*}
\renewcommand{\arraystretch}{1.55} 
\setlength{\tabcolsep}{14pt}
\centering
\begin{tabular}{lccccc}
\hline \hline
Parameter & Prior & Visit 1 & Visit 2 & Combined  \\ \hline
$\log (\mathrm{P}_\mathrm{ref}/\mathrm{bar})$ & U(-6, 0) & $-2.07_{-0.13}^{+0.10}$ &  $-1.65_{-0.08}^{+0.06}$ &  $-1.85_{-0.07}^{+0.07}$ \\
T$_\mathrm{iso,\,E}$ (K) & U(450, 2000) &  $1512_{-73}^{+74}$ & $1558_{-65}^{+75}$ &  $1545_{-54}^{+54}$ \\ 
T$_\mathrm{iso,\,M}$ (K) & U(450, 2000) &  $960_{-110}^{+130}$ & $740_{-110}^{+130}$ & $881_{-94}^{+100}$ \\ 
$\log (\mathrm{X}_{\mathrm{H}_2\mathrm{O}})$ & U(-12, -0.3) & $-4.27_{-0.13}^{+0.14}$ & $-4.53_{-0.11}^{+0.12}$ & $-4.44_{-0.09}^{+0.10}$ \\
$\log (\mathrm{X}_{\mathrm{CO}_2})$ & U(-12, -0.3) & $-7.26_{-0.27}^{+0.25}$ & $-8.62_{-1.37}^{+0.54}$ & $-7.75_{-0.22}^{+0.19}$ \\
$\log (\mathrm{X}_{\mathrm{H}_2\mathrm{S}})$ & U(-12, -0.3) & $-4.83_{-0.22}^{+0.20}$ & $< -5.9$ & $-5.33_{-0.25}^{+0.18}$ \\
$\log (\mathrm{X}_{\mathrm{CO}})$ & U(-12, -0.3) & $< -5.3$ & $< -5.3$ & $< -5.7$ \\
$\log (\mathrm{P}_\mathrm{deck,\,E}/\mathrm{bar})$ & U(-3, 2) & $0.55_{-0.98}^{+0.97}$ & $0.45_{-0.95}^{+1.00}$ &  $0.44_{-0.89}^{+1.04}$ \\
$\log (\mathrm{P}_\mathrm{deck,\,M}/\mathrm{bar})$ & U(-3, 2) & $0.2_{-1.2}^{+1.2}$ & $0.45_{-0.96}^{+0.96}$ & $0.4_{-1.0}^{+1.0}$  \\
$\log(r_{\mathrm{m,\,E}}/\mu \mathrm{m})$ & U(-3, 1) & $-1.60_{-0.92}^{+1.38}$ & $-1.65_{-0.90}^{+1.35}$ & $-1.61_{-0.93}^{+1.43}$ \\
$\log(r_{\mathrm{m,\,M}}/\mu \mathrm{m})$ & U(-3, 1) & $-1.34_{-0.28}^{+0.27}$ & $-1.38_{-0.27}^{+0.27}$ & $-1.23_{-0.27}^{+0.22}$ \\
$\log(n_{\mathrm{max,\,MgSiO_3,\,E}})$ & U(0, 25) & $6.8_{-4.6}^{+6.8}$ & $6.7_{-4.4}^{+6.4}$ & $6.6_{-4.5}^{+6.7}$ \\
$\log(n_{\mathrm{max,\,MgSiO_3,\,M}})$ & U(0, 25) & $11.8_{-2.3}^{+2.6}$ & $11.6_{-2.0}^{+2.2}$ & $11.0_{-1.9}^{+2.4}$ \\
$f_\mathrm{E}$ & U(0, 1) & $0.36_{-0.26}^{+0.38}$ & $0.34_{-0.24}^{+0.37}$ & $0.38_{-0.28}^{+0.39}$ \\
$f_\mathrm{M}$ & U(0, 1) & $0.68_{-0.14}^{+0.16}$ & $0.75_{-0.15}^{+0.14}$ & $0.72_{-0.14}^{+0.15}$ \\ \hline
\end{tabular}
\vspace{1mm}
\caption{Retrieved atmospheric parameters for the limb-resolved analysis of Kepler-12~b, as described in Sections~\ref{sec:asymmetric} and \ref{sec:time-resolved}.
}
\label{tab:asymmetric_atm_params}
\end{table*}


\section{Individual-visit morning and evening spectra} \label{app:individual-visit}

\begin{figure*}
\centering
	\includegraphics[width=1\textwidth]{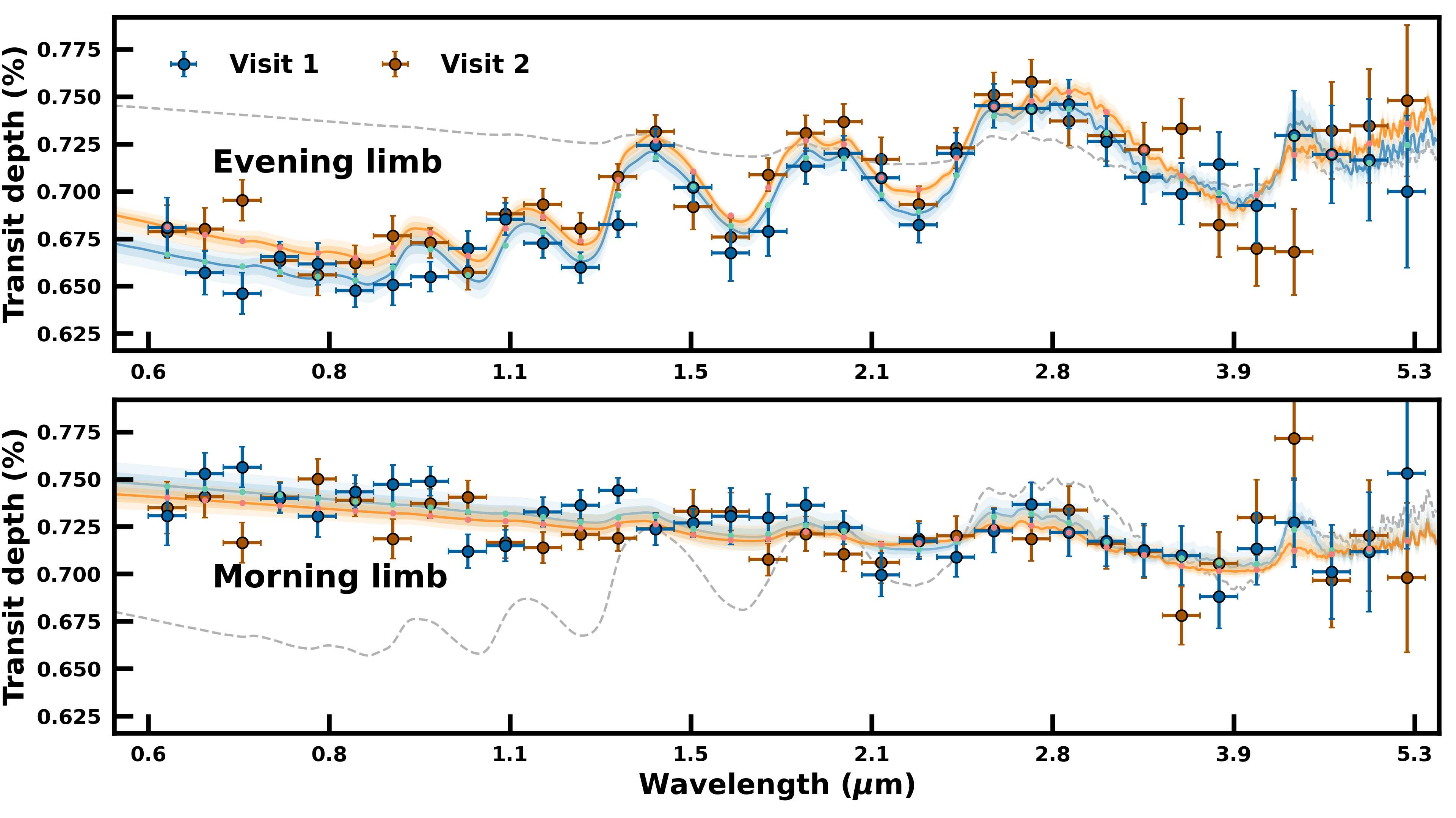}
    \vspace{-5mm}
    \caption{Same as Figure~\ref{fig:asymmetric_spectrum}, but showing both visits separately. The data is binned to $R = 15$ for clarity. The blue and orange data points and models correspond to visit 1 and 2, respectability. The morning model spectra have been offset $+67$~ppm while the evening model spectra have been offset $-67$~ppm for visual clarity. }
    \label{fig:asymmetric_spectrum_time} 
\end{figure*}

Figure~\ref{fig:asymmetric_spectrum_time} shows the morning and evening spectra of the two visits, along with retrieved atmospheric models. The retrieved parameters for both visits are given in Table~\ref{tab:asymmetric_atm_params}. Note that the time variability of these spectra is more uncertain than that of the limb-averaged spectra, as discussed in Sections~\ref{sec:white_time} and \ref{sec:time-resolved}.

\bibliography{ms, references}{}
\bibliographystyle{aasjournalv7}



\end{document}